\documentclass[a4paper,fleqn,usenatbib]{mnras}
\usepackage{newtxtext,newtxmath}

\usepackage[T1]{fontenc}
\usepackage{ae,aecompl}

\usepackage{graphicx}	
\usepackage{amsmath}	
\usepackage{amssymb}	

\title[QSAGE - II: quasar environments at $z=1-2$]{Quasar Sightline and Galaxy Evolution (QSAGE) survey - II. Galaxy overdensities around UV luminous quasars at $z=1-2$}

\author[J. P. Stott et al.]{
J. P. Stott,$^{1}$\thanks{E-mail: j.p.stott@lancaster.ac.uk (JPS)}
R. M. Bielby,$^{2}$
F. Cullen,$^{3}$
J. N. Burchett,$^{4}$
N. Tejos,$^{5}$
\newauthor
M. Fumagalli,$^{2,6,7}$
R. A. Crain,$^{8}$
S. L. Morris,$^{2}$
N. Amos,$^{1,9}$
R. G. Bower,$^{7}$
\newauthor
J. X. Prochaska.$^{4}$
\\
$^{1}$Department of Physics, Lancaster University, Lancaster LA1 4YB, UK\\
$^{2}$Centre for Extragalactic Astronomy, Durham University, South Road, Durham, DH1 3LE, UK\\
$^{3}$Institute for Astronomy, University of Edinburgh, Royal Observatory, Edinburgh EH9 3HJ, UK\\
$^{4}$UCO/Lick Observatory, University of California, Santa Cruz, CA, USA\\
$^{5}$Instituto de F\'{i}sica, Pontificia Universidad Cat\'{o}lica de Valpara\'{i}so, Casilla 4059, Valpara\'{i}so, Chile\\
$^{6}$Dipartimento di Fisica G. Occhialini, Universit\`a degli Studi di Milano Bicocca, Piazza della Scienza 3, 20126 Milano, Italy \\
$^{7}$Institute for Computational Cosmology, Durham University, South Road, Durham, DH1 3LE, UK \\
$^{8}$Astrophysics Research Institute, Liverpool John Moores University, 146 Brownlow Hill, Liverpool L3 5RF, UK\\
$^{9}$Isaac Newton Group of Telescopes, Apartado de Correos 321, E-38700 Santa Cruz de La Palma, Spain}

\date{Accepted XXX. Received YYY; in original form ZZZ}

\pubyear{2019}

\begin{document}
\label{firstpage}
\pagerange{\pageref{firstpage}--\pageref{lastpage}}
\maketitle

\begin{abstract}

We demonstrate that the UV brightest quasars at $z=1-2$ live in overdense environments. This is based on an analysis of deep {\it Hubble Space Telescope} WFC3 G141 grism spectroscopy of the galaxies along the lines-of-sight to UV luminous quasars in the redshift range $z=1-2$. This constitutes some of the deepest grism spectroscopy performed by WFC3, with 4 roll angles spread over a year of observations to mitigate the effect of overlapping spectra. Of the 12 quasar fields studied, 8 display evidence for a galaxy overdensity at the redshift of the quasar. One of the overdensities, PG0117+213 at $z=1.50$, has potentially 36 spectroscopically confirmed members, consisting of 19 with secure redshifts and 17 with single-line redshifts, within a cylinder of radius $\sim700$\,kpc. Its halo mass is estimated to be $\log (M/M_\odot)=14.7$. This demonstrates that spectroscopic and narrow-band observations around distant UV bright quasars may be an excellent route for discovering protoclusters. Our findings agree with previous hints from statistical observations of the quasar population and theoretical works, as feedback regulated black hole growth predicts a correlation between quasar luminosity and halo mass. We also present the high signal-to-noise rest-frame optical spectral and photometric properties of the quasars themselves.

\end{abstract}




\section{Introduction}
\label{sec:intro}

Detecting galaxy clusters and protoclusters beyond $z\sim1$ is challenging but important, as such overdensities are probes of both the accelerated galaxy evolution in dense environments and can provide constraints on cosmological parameters. {Modern cluster cosmology generally achieves this via studies of cluster counts (e.g. \citealt{vikhlinin2009,mantz2010,dehaan2016}) or the mass fraction of the intracluster medium (ICM) in the most massive systems (e.g. \citealt{allen2008,ettori2009,mantz2014}).} The dominant conventional methods for detecting clusters at $z<1$ are via the X-ray emission of the hot {ICM (e.g.} \citealt{ebeling1998,bohringer2004,finoguenov2007,mehrtens2012}), the Sunyaev-Zel'dovich (SZ) effect \citep{sunyaev1972} due to the scattering of cosmic microwave background photons by the ICM (e.g. \citealt{staniszewski2009,vanderlinde2010,planck2014,bleem2015,hilton2018}), and searching for overdensities of passive red galaxies in photometric data (e.g. \citealt{gladders2005,koester2007,rykoff2014,chan2019}). The flux limited nature of X-ray surveys means it is difficult to find all but the most massive clusters at $z>1$ in significant numbers and in the protocluster phase there may not yet be a mature ICM to provide significant emission. The SZ effect, being a spectral distortion of the microwave background, is in principle redshift independent but still relies on the presence of a mature ICM for the signal so is again not the right tool for selecting less evolved systems. The problem with selecting clusters via overdensities of red galaxies at $z>1$, is that cluster galaxies are likely to be building up their mass through star formation and are therefore no longer homogenous in colour \citep{hilton2010,tran2010}. 

One technique for discovering clusters and protoclusters at $z>1$, is to search for galaxy overdensities around radio galaxies, which are often the most massive galaxies for their epoch and are therefore likely to live in a clustered environment (e.g. \citealt{lefevre1996,best1998,pentericci2000,kurk2004,venemans2007,hatch2011,wylezalek2013,castignani2014,husband2016}). This provides a connection and potential evolutionary link to the brightest cluster galaxies (BCGs) found in the low redshift Universe, which are the most massive galaxies and often radio-loud active galactic nuclei (AGN) themselves (e.g. \citealt{best2007,stott2012}). However, it is important to note that not all distant radio galaxies are found in clustered environments \citep{hatch2011,wylezalek2013}.

In terms of quasars, \cite{wylezalek2013} find that the majority (55\%) of their radio-loud galaxies were in $>2\sigma_\delta$ overdensities and there was no significant difference between the environments of radio-loud galaxies or radio-loud quasars. However, radio-quiet AGN such as optical, infrared or X-ray selected quasars are typically found in lower mass galaxies and in less clustered environments, with typical halo masses $10^{12}-10^{13}$\,M$_\odot$ (\citealt{kauffmann2008,hickox2009,ross2009,shen2009,geach2019}). The environmental difference between radio-loud and radio-quiet AGN may be explained by `jet confinement', in which the denser intergalactic/intracluster medium found in overdensities enhances radiation losses making the radio jets brighter \citep{barthel1996}. In this case an AGN is more likely to present as radio-loud if in a dense environment. Because of the comparatively low halo masses of optically selected quasars, protocluster searches are usually conducted around radio galaxies instead. However, there are some radio-quiet quasars in dense environments (e.g. \citealt{haines2001}) and there has been some success in finding overdensities using pairs of quasars as tracers \citep{boris2007}.

While it appears that in general, optically selected quasars live in lower density environments than radio galaxies, there is some evidence that the most optically luminous reside in more massive dark matter halos. In their statistical study of halo mass based on the gravitational deflection of the cosmic microwave background, \cite{geach2019} find that the average halo mass of an optically selected quasar is $\log (M_h/h^{-1} M_\odot) =12.6\pm0.2$ but those with an absolute $i-$band magnitude $M_i\lesssim-26$ are in $M_h \approx10^{13} M_\odot h^{-1}$ haloes. This is in agreement with the clustering observations of \cite{shen2009}, who find that the 10\% most optically luminous quasars are more strongly clustered. However, there is also observational evidence that UV luminous quasars do not live in high mass halos. For example, a galaxy-quasar cross-correlation analysis of a sample of `hyperluminous' quasars at $z\sim3$, concludes that they live in modest $\log (M_h/h^{-1} M_\odot) =12.3\pm0.5$ haloes \citep{trainor2012}. In addition to this, no correlation is found between the locations of bright quasars and a catalogue of overdensities at $z=4$ \citep{uchiyama2018}.

Despite the uncertain observational picture presented above, there is a theoretical expectation that the {UV} brightest quasars should live in more massive dark matter halos, as the feedback regulated growth of black holes should result in a relation between black hole mass and halo mass e.g. $M_{bh}\propto M_h^{5/3}$ (\citealt{silk1998,wyithe2003}). Assuming quasars accrete at an approximately fixed fraction fraction of the Eddingtion limit on average, then the quasar luminosity $L\propto M_h^{5/3}$. Correlations between $M_{bh}$ and $M_h$ are indeed seen in hydrodynamic cosmological simulations of galaxy evolution that employ black hole feedback \citep{booth2010,mcalpine2017}. One can also think of this as a consequence of the stellar mass -- halo mass relation (e.g. \citealt{conroy2009}) and the black hole mass -- galaxy mass relation (e.g. \citealt{magorrian1998,ferrarese2000,gebhardt2000}). 

In this paper we test whether UV/optically luminous quasars do indeed trace high density environments using data from the Quasar Sightline And Galaxy Evolution (QSAGE) survey. The QSAGE survey \citep{bielby2019} was designed to perform deep {\it Hubble Space Telescope} (HST) Wide-Field Camera 3 (WFC3) grism observations centred on 12 quasars with existing high-quality HST UV spectra from the Space Telescope Imaging Spectrograph (STIS) and the Cosmic Origins Spectrograph (COS). The primary purpose of QSAGE is to obtain hundreds of galaxy redshifts in order to associate absorption features in the UV spectra with their host galaxies' circum-galactic medium (CGM). This information is being used to constrain models of galaxy fuelling and feedback (e.g. \citealt{bielby2017,bielby2019}).

The paper begins in \S\ref{sec:data} with a description of the QSAGE sample and data reduction, including the spectral and photometric properties of galaxies and the quasars themselves. In \S\ref{sec:ana} we present the overdensity analysis and results, and look for dependencies between environment and both quasar and galaxy properties. Our findings are discussed in \S\ref{sec:disc} and conclusions drawn in \S\ref{sec:conc}.

We adopt a cosmology with $\Omega_{\Lambda}$\,=\,0.7, $\Omega_{m}$\,=\,0.3, and H$_{0}$\,=\,70\,km\,s$^{-1}$\,Mpc$^{-1}$. All quoted magnitudes are on the AB system and we use a \cite{chabrier2003} IMF throughout.

\section{Sample and Data}
\label{sec:data}

\subsection{WFC3 Data Reduction}
\label{sec:wfc3}

The QSAGE survey targets 12 $z=1.2-2.4$ quasars that have existing high quality UV spectra, with the aim of obtaining the redshifts and properties of the galaxy populations along their lines-of-sight (HST Cycle 24 Large Program 14594; PIs: R. Bielby, J. P. Stott). The primary purpose of QSAGE is to perform CGM studies in the redshift range $0.6\lesssim z \lesssim 2.4$ but there is a wealth of additional galaxy evolution science that can be done with the data. The full survey consists of 96 orbits with WFC3 in imaging and grism mode (i.e., 8 orbits per quasar sightline).  A detailed description of the spectral and imaging data and its reduction is provided in \cite{bielby2019}, however we briefly describe the key points here.

The HST observations for an individual quasar field consist of 8 grism exposures and supporting near-infrared (NIR) imaging, in order to identify the coordinates of the sources in the grism data. The observations for an individual field are divided into four visits spread out over the year-long HST cycle in order to ensure that each visit is at a different roll angle. This is key because the depth of the grism data requires multiple roll angles to minimise the contamination from the overlapping spectra of neighbouring sources. A visit comprises two orbits, with each orbit scheduled to acquire the following: a single $\sim250$s F140W image; 2 G141 grism exposures of $\sim1000$s; and a single $\sim250$s F160W image. In practice, the exposure times were tweaked to optimally fill a given orbit.

The imaging and grism data were reduced using the \textsc{Grizli} software package (Brammer, in prep.\footnote{https://github.com/gbrammer/grizli}), which is designed for the reduction and analysis of slitless spectroscopic datasets. \textsc{Grizli} builds on the previous \textsc{aXe} \citep{kummel2009} and \textsc{threedhst} (\citealt{brammer2012}; \citealt{momcheva2016}) software. The F140W and F160W imaging was reduced using \textsc{TweakReg} and then \textsc{AstroDrizzle}. The individual images of a field were combined in a median stack using {\sc SWarp} \citep{bertin2002}. The photometric zeropoints were calculated from the image header information ($\rm F140W_{ZP}=26.45$ and $\rm F160W_{ZP}=25.95$). The photometric catalogues were produced using {\sc SExtractor} \citep{bertin1996} with the F140W catalogue used as the input source coordinates for the G141 grism data.

\begin{figure}
		\includegraphics[width=\columnwidth]{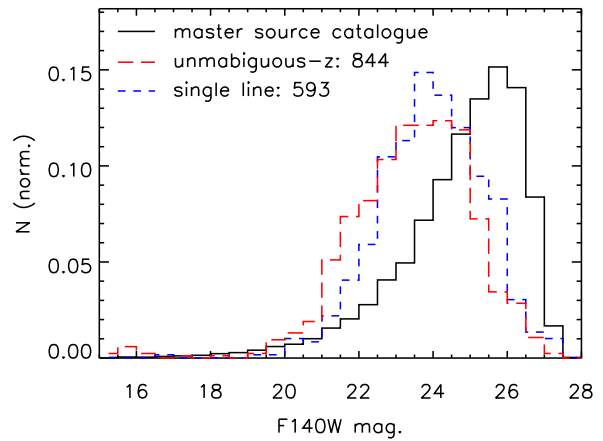}

	     \caption{The F140W magnitude distribution of the QSAGE WFC3 galaxy sample. These are \textsc{SExtractor} $\rm MAG\_AUTO$ magnitudes. The black solid line represents all of the WFC3 detections. The long dashed red line represents the galaxies with unambiguous redshifts. The dashed blue line represents the galaxies with single line detections and therefore ambiguous redshifts. All distributions are normalised to a total area of 1.}

 \label{fig:maghistall}
\end{figure}

The approximate 50\% completeness values of the F140W and F160W observations are 26.3 mag and 26.1 mag respectively (\textsc{SExtractor} $\rm MAG\_AUTO$). These limits are calculated by fitting a straight line to the logarithmic number counts of the magnitude distribution up to its peak then extrapolating this fit beyond the peak and comparing it with the observed number counts {\bf(see Fig. \ref{fig:maglim})}. The F140W magnitude distribution for the QSAGE sources is presented in Fig. \ref{fig:maghistall}.

The F140W catalogues and segmentation images generated by \textsc{SExtractor} were used to extract the spectrum of each object with \textsc{Grizli}. Each of the G141 exposures is first divided by the F140W flat field. An initial background was then subtracted using the `master sky' images from \citet{brammer2015}.
These master sky images account for the variation in background structure due to variations in zodiacal continuum, scattered light, and He emission across the sky \citep{momcheva2016}. The residuals (typically $0.5-1 \%$ of the initial background levels) were then removed by subtracting average values of the sky pixels in each column. Finally 1D and 2D spectra were extracted from the background-subtracted grism images at the individual exposure level, which, for the \textsc{QSAGE} observations, resulted in 16 individual spectra per object (i.e. 2 spectra per orbit). The advantage of having individual spectra spread out over 4 visits at different times in the HST cycle is that they were taken at four different roll angles. This substantially mitigates the spectral contamination due to neighbouring sources. The contamination for each individual spectrum was calculated using \textsc{Grizli}, which was used to mask heavily contaminated pixels when generating the final stacked spectrum for each source. To aid with visual inspection of the spectra (see \S\ref{sec:spec}; Fig. 3 of \citealt{bielby2019}), a stack was created for each of the four visits (roll angles).

\subsection{Spectral fitting}
\label{sec:spec}

Grism spectra were extracted for a total of 13223 objects across the 12 QSAGE fields. The first step in our analysis was to estimate the emission line redshifts of the galaxies by allowing multiple Gaussians to be fit (with all parameters free, including wavelength) to each spectrum by randomly placing 10 of them throughout the wavelength range in order to search for emission line candidates. This was done iteratively with the first fitted line subtracted from the data before the next fit is attempted, to ensure subsequent fits found different lines. The wavelength range of the WFC3 G141 grism runs from approximately 1075 nm to 1700 nm. The prominent lines and their in-vacuo wavelengths seen in the spectra of star-forming galaxies and AGN in the rest-frame optical range are: [OII] (372.709\,nm, 372.988\,nm); [OIII] (496.030\,nm, 500.824\,nm); and H$\alpha$ (656.461\,nm). The weaker lines that are often still bright enough to easily detect are H$\beta$ (486.272\,nm) and [SII] (671.829\,nm, 673.268\,nm).  An estimated redshift solution was found by taking the wavelength ratios of significant ($\rm S/N>3$) Gaussian fits to determine whether they corresponded to the wavelength ratios of $\lambda_{\rm [OII]}/\lambda_{\rm [OIII]}\approx0.744$ or $\lambda_{\rm [OIII]}/\lambda_{\rm [H\alpha]}\approx0.763$. At this stage multiple redshift solutions were allowed for each galaxy. All $\sim13$K spectra were then visually inspected alongside the spectra and contaminations for each of the 4 visits (roll angles), and a postage stamp F140W image of the galaxy. This was to either choose the correct redshift solution, provide a new one based on this visual inspection or dismiss the source as a poor solution/spectrum. During this process the spectra were also given a quality flag from $1 - 4$ with 1 being a poor quality noise-dominated spectra, 2 being low signal-to-noise spectra but with potential spectral lines ($\rm S/N<3$), 3 being good quality spectra (lines with $\rm S/N>3$) that are suitable for redshift calculations and 4 being those with higher signal-to-noise ($\rm S/N>10$).  At this point low-$z$ absorption line galaxies, stars, artefacts and potentially interesting objects such as candidate high redshift galaxies and cool brown dwarfs were also categorised. 

\begin{figure}
		\includegraphics[width=\columnwidth]{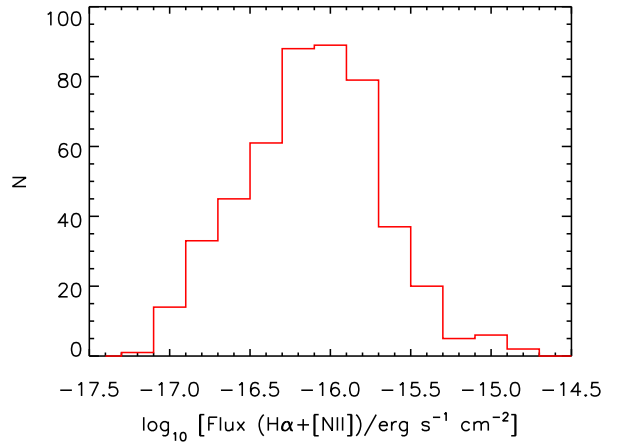}
    \caption{The flux distribution of the blended H$\alpha$+[NII] lines for the unambiguous-$z$ galaxy sample.}
	    \label{fig:fluxhist}
\end{figure}

For the redshift and emission-line analysis in the remainder of the paper we only consider quality 3 and 4 galaxy spectra {(example spectra are presented in Fig. \ref{fig:exspec})}. The emission-line fluxes are calculated by fitting Gaussian profiles and a local linear continuum. The low resolution of the WFC3 G141 grism ($R=130$ at 1400\,nm) means that the [OII], [OIII] and [SII] doublets, and H$\alpha$ and [NII] are unresolved and blended as single lines. For the [OII] we fit a single Gaussian centred on the mid-point of the doublet (372.849\,nm). For the [OIII] doublet and H$\beta$ we fit a triplet of Gaussians, as while the [OIII] doublet lines are blended together, they are marginally resolved to the extent that one can observe a distinct asymmetric shape due to the lower flux of the [OIII]\,496.030\,nm line compared with the 500.824\,nm line. The H$\beta$ line is included in the triplet with the [OIII] lines as the low resolution may lead to blending in the wings of these lines. When fitting we assume a fixed flux ratio in the [OIII] doublet of 2.98 \citep{storey2000}. Again because of the grism resolution, we fit the H$\alpha$ and [SII] lines as a doublet of two Gaussians. The H$\alpha$ line is fully blended into a single Gaussian with the adjacent [NII] lines (654.986\,nm, 658.527\,nm). To attempt to extract the true H$\alpha$ flux, we assume that the nebula gas has a solar abundance ($12+\log\rm(O/H)=8.66$, \citealt{asplund2004}) and thus assuming the linear relation of \cite{pettini2004}, the ratio of [NII]\,658.527/H$\alpha$ is fixed at 0.379. We note however, that solar metallicity is likely too high for galaxies at $1<z<2$ \citep{maiolino2008}. {In the absence of individual metallicity information for the galaxies, the effect of this is to potentially underestimate the H$\alpha$ flux and therefore star formation rate (SFR) of the sample. If there is a significant difference in metallicity between the overdensity and field, this could potentially affect the comparison of SFR with environment presented in \S\ref{sec:dep} }. The flux ratio of the redder 658.527\,nm is fixed to be 2.95 greater than that of the fainter blue [NII]\rm654.986\,nm line, as is the theoretical expectation. The [SII] doublet is completely blended and so we treat it as a single Gaussian centred on the average wavelength of 672.548\,nm. To obtain the completeness limit of the spectroscopy, the flux distribution of the Gaussian fits to the blended H$\alpha$ and [NII] lines is presented in Fig. \ref{fig:fluxhist}. This has a median flux of $7.9\pm 0.8\times10^{-17}$\,erg\,s$^{-1}$\,cm$^{-2}$ and an approximate 50\% completeness of $7.9\times10^{-17}$\,erg\,s$^{-1}$\,cm$^{-2}$.   

\begin{figure}
		\includegraphics[width=\columnwidth]{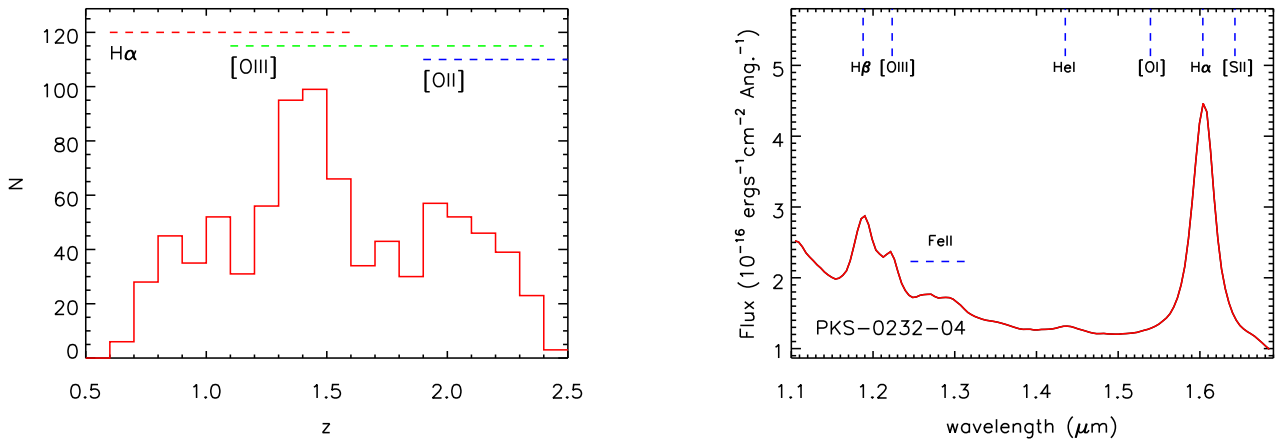}
    \caption{The redshift distribution of the unambiguous-$z$ galaxy sample. The structure seen mainly reflects the ease of obtaining redshifts in windows containing multiple bright lines, i.e. [OIII] and H$\alpha$ at $z\sim1.1 - 1.6$, and [OII] and [OIII] at $z\sim1.9 - 2.4$. The dashed horizontal lines above the distribution represent the redshift ranges where each of these lines can be detected.}
	    \label{fig:zhistall}
\end{figure}

The redshifts we can access for the majority of the QSAGE galaxy population are: $z\sim1.9 - 3.6$ ([OII]), $z\sim1.2 - 2.5$ (H$\beta$), $z\sim1.1 - 2.4$ ([OIII]) and $z\sim0.6 - 1.6$ (H$\alpha$). There is therefore, continuous coverage in redshift from $z=0.64 - 3.6$. Within this, there are ranges in redshift where the spectroscopic redshift is unambiguous, as multiple bright lines are present within the grism's wavelength window. These multiple line windows are [OIII] and H$\alpha$ at $z\sim1.1 - 1.6$, and [OII] and [OIII] at $z\sim1.9 - 2.4$. Where the, generally weaker, secondary line is visible it is also possible to obtain unambiguous redshifts from H$\beta$ and [OIII] at $z\sim1.2 - 2.4$ and H$\alpha$ and [SII] at $z\sim0.6 - 1.5$. Finally, in high signal to noise cases, the asymmetry in the blended [OIII] (496.030\,nm, 500.824\,nm) can also provide an unambiguous redshift from $z=1.1 - 2.4$. In cases of single and therefore ambiguous lines we do not assign a redshift, although we show the result of assuming a line identity in \S\ref{sec:env}. In total, we obtain good spectra for 1437 galaxies. Of these, 844 have unambiguous redshifts with the remaining 593 being ambiguous single-line detections. For the remainder of the paper, we will refer to the galaxies with unambiguous redshifts as the {\it unambiguous-$z$} sample. The distribution of the unambiguous-$z$ sample redshifts is plotted in Fig. \ref{fig:zhistall}. We note the peak in this graph at $z=1.2-1.6$, which reflects the ease of obtaining unambiguous redshifts in that window as discussed above. In order to break the single-line degeneracy and produce a more complete set of QSAGE redshifts, spectral energy distribution (SED) fitting will be performed and presented in Bielby et al., in prep.

\subsection{Quasar properties}
\label{sec:qso}

\begin{figure}
		\centering
		\includegraphics[width=\columnwidth, trim=0 0 0 0, clip=true]{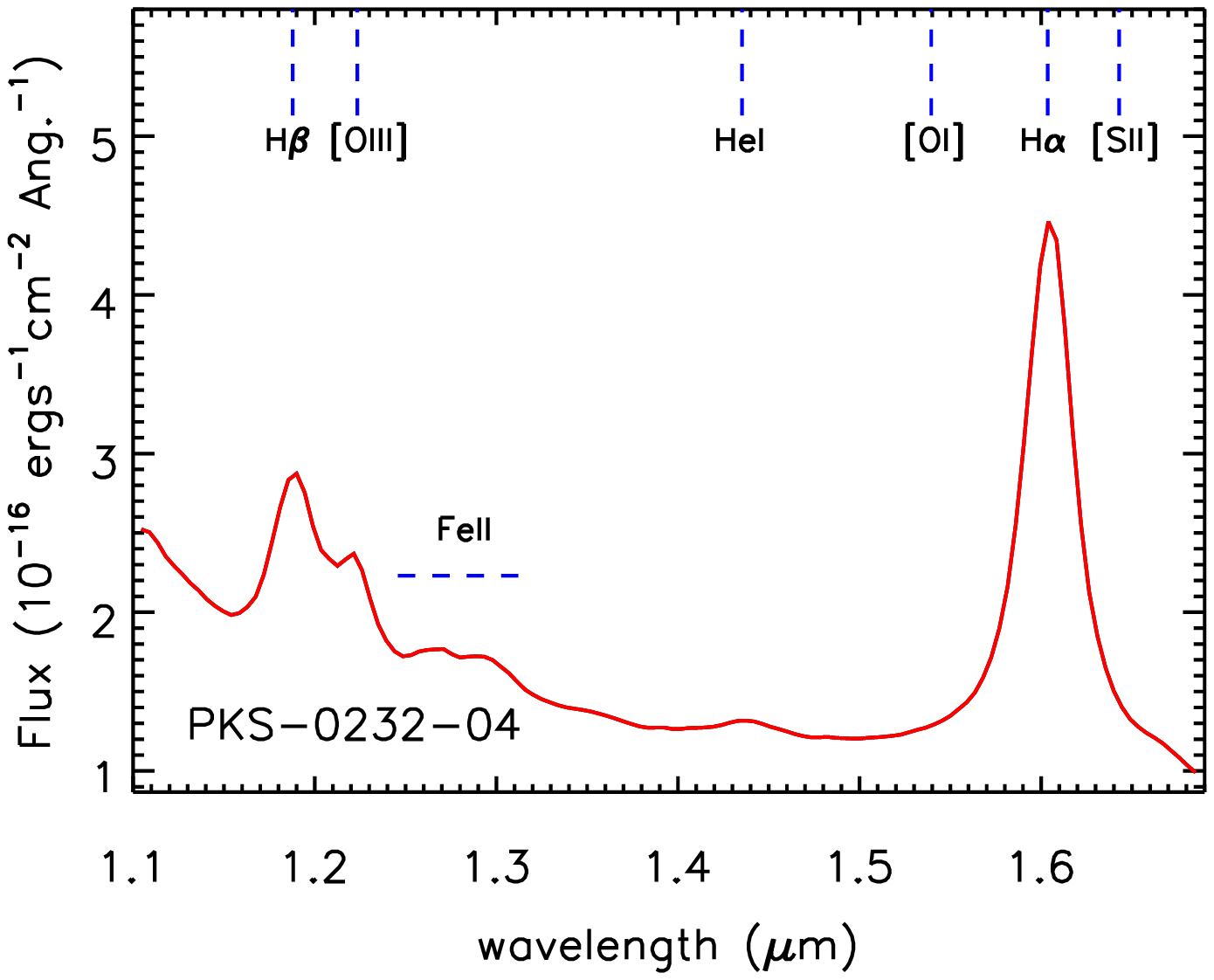}
		\includegraphics[width=\columnwidth, trim=0 0 0 0, clip=true]{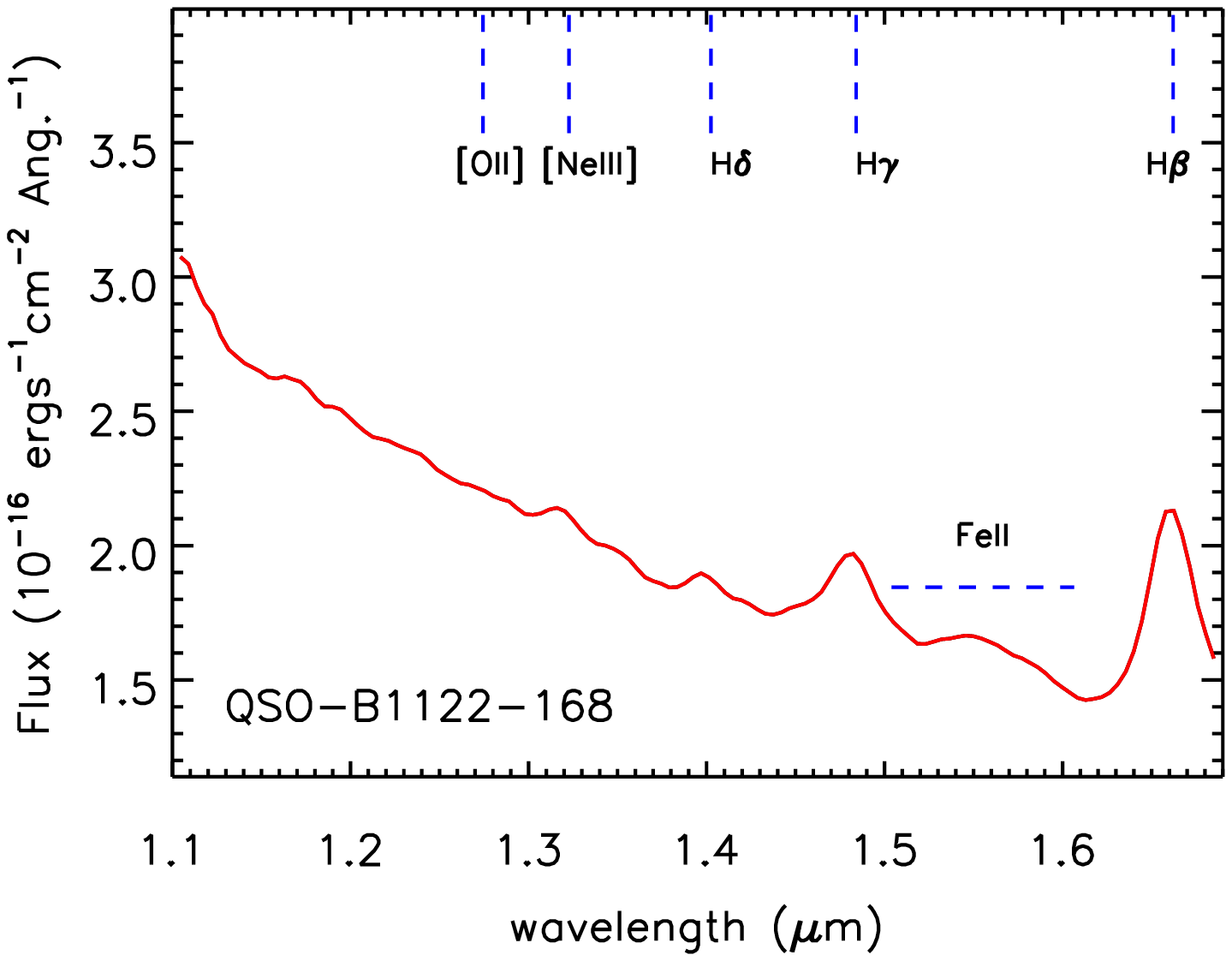}
    \caption[]{{\it Upper}: PKS-0232-04 at $z=1.4428$ is a typical example of the extremely high signal-to-noise QSAGE WFC3 quasar spectra. The spectrum contains the bright emission lines H$\beta$, [OIII] and H$\alpha$ (dominating over the [SII] emission), and the weaker HeI line. The FeII complex of multiplets at $5100-5400$\AA \, are also present. {\it Lower}: The spectrum of the highest redshift quasar in our sample is QSO-B1122-168 at $z=2.4181$, which contains the emission lines H$\beta$, H$\gamma$, H$\delta$, [NeIII] and the FeII complex of multiplets at $4400-4700$\AA. The lack of [OII] emission is likely due to the radiation from the quasar causing a high ionisation state, which means that most of the oxygen emission is via [OIII] instead.}
	    \label{fig:qsospec}
\end{figure}

The 12 central quasars of the QSAGE sample are presented with their redshifts, photometry and emission line fluxes in Table \ref{tab:sample}. All of the quasar spectra with the necessary wavelength coverage also show the presence of H$\gamma$, H$\delta$, [OI], [NeIII] and the FeII multiplet complexes at $4400-4700$\AA\, and $5100-5400$\AA \, (see \citealt{phillips1978} and \citealt{zhou2002}). Example WFC3 quasar spectra are shown in Fig. \ref{fig:qsospec}. A column for [OII] flux is not included in Table \ref{tab:sample} as QSO-B1122-168 is the only quasar with this line in its redshift range and despite a very high signal-to-noise spectrum, it is not detected (see Fig. \ref{fig:qsospec}). This is likely due to the radiation from the quasar causing a high ionisation state, which means that most of the oxygen emission is via [OIII] instead (although this is out of the wavelength window for this quasar). Any [OII] emission due to star formation in the host is likely to be completely overwhelmed by the continuum emission of the quasar. 

\begin{figure}
		\centering
		\includegraphics[width=0.9\columnwidth, trim=0 0 0 0, clip=true]{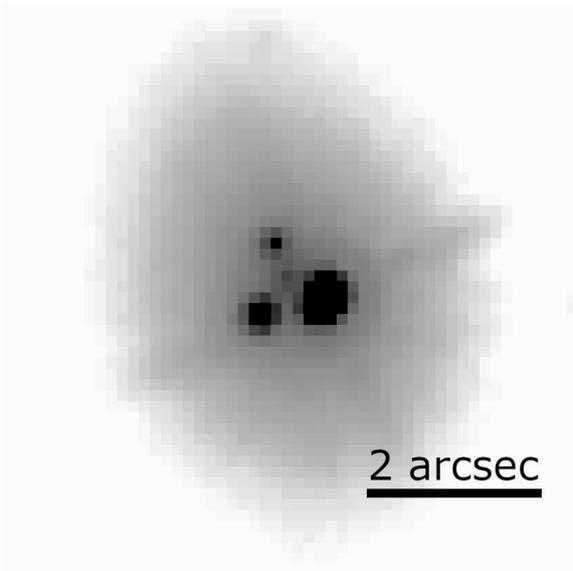}
    \caption{This F140W image is of the multiply lensed quasar QSO-B0810+2554.  A logarithmic image scale is used to display the lensed components. Our image shows it as triply lensed although it appears the larger of the three components separates into 2 components making a quadruple \citep{reimars2002}. This may indicate that the apparent UV luminosity of this quasar is enhanced due to lensing.}
	    \label{fig:qsolens}
\end{figure}

QSO-B0810+2554 is a known quadruply lensed quasar \citep{reimars2002}, which is potentially interesting for cosmology via gravitational lens time delays (e.g. \citealt{congdon2010,wong2019}). The QSAGE F140W image of this lens is displayed in Fig. \ref{fig:qsolens}. We find no evidence for lensing in the remaining QSAGE quasars.

\begin{table*}
	\footnotesize
	\centering
	\caption{The QSAGE central quasar sample. The right ascension (R.A.), declination (Dec.) and redshift ($z$) are those we derive from our WFC3 imaging and spectroscopy, which in some cases are an update to the literature values. The emission line fluxes are in units of $10^{-14}$\,erg\,s$^{-1}$\,cm$^{-2}$. The apparent $B$-band magnitudes, $m_B$, are from the United States Naval Observatory (USNO) A2 catalogue (\citealt{monet1998}).}
	\label{tab:sample}
	\begin{tabular}{lccccccccc} 
		\hline
		Quasar & R.A. & Dec. & $z$ & $m_B$ & F140W & F160W & $f_{\rm H\beta}$ & $f_{\rm [OIII]}$ & $f_{\rm H\alpha}$   \\
		\hline

PG0117+213 & 01:20:17.3 & +21:33:46.2 & 1.5041 & 15.5 & 15.8 & 15.6 & 1.69$\pm$0.29 & 0.70$\pm$0.19 & 12.66$\pm$1.03  \\
PKS-0232-04 & 02:35:07.3 & -04:02:05.3 & 1.4428 & 16.3 & 16.3 & 16.1 & 2.51$\pm$0.29 & 1.03$\pm$0.19 & 10.99$\pm$0.44  \\
HE0515-4414 & 05:17:07.6 & -44:10:55.6 & 1.7332 & 15.1 & 14.6 & 14.6 & 4.38$\pm$0.35 & 1.28$\pm$0.24 & ...  \\
2QSO-B0747+4259 & 07:50:54.6 & +42:52:19.3 & 1.9151 & 15.6 & 15.5 & 15.5 & 2.78$\pm$0.17 & 0.74$\pm$0.11 & ...  \\
QSO-B0810+2554 & 08:13:31.3 & +25:45:03.1 & 1.5103 & 15.5 & 15.2 & 14.9 & 5.25$\pm$0.53 & 0.86$\pm$0.32 & 33.45$\pm$1.87  \\
QSO-J1019+2745 & 10:19:56.6 & +27:44:01.7 & 1.9298 & 15.6 & 16.0 & 15.9 & 2.66$\pm$0.16 & 0.99$\pm$0.11 & ...  \\
QSO-B1122-168 & 11:24:42.9 & -17:05:17.4 & 2.4181 & 16.5 & 16.1 & 16.1 & 2.48$\pm$0.17 & ... & ...  \\
QSO-J1130-1449 & 11:30:07.1 & -14:49:27.4 & 1.1896 & 16.1 & 17.2 & 23.9 & ... & ... & 5.71$\pm$0.13  \\
LBQS-1435-0134 & 14:37:48.3 & -01:47:10.8 & 1.3061 & 15.1 & 15.3 & 15.1 & 5.42$\pm$0.90 & 3.76$\pm$0.68 & 18.28$\pm$0.55  \\
QSO-B1521+1009 & 15:24:24.5 & +09:58:29.1 & 1.3266 & 14.7 & 15.8 & 15.7 & 4.48$\pm$1.17 & 1.31$\pm$0.75 & 10.43$\pm$0.36  \\
QSO-B1630+3744 & 16:32:01.1 & +37:37:50.0 & 1.4787 & 16.0 & 15.9 & 15.7 & 1.98$\pm$0.27 & 0.54$\pm$0.17 & 10.61$\pm$0.51  \\
QSO-B1634+7037 & 16:34:29.0 & +70:31:32.4 & 1.3319 & 14.9 & 14.3 & 14.1 & 7.90$\pm$1.80 & 8.32$\pm$1.36 & 55.32$\pm$1.75  \\

		\hline
	\end{tabular}
\end{table*}

\subsection{Star formation rates}

\begin{figure}
		\includegraphics[width=\columnwidth]{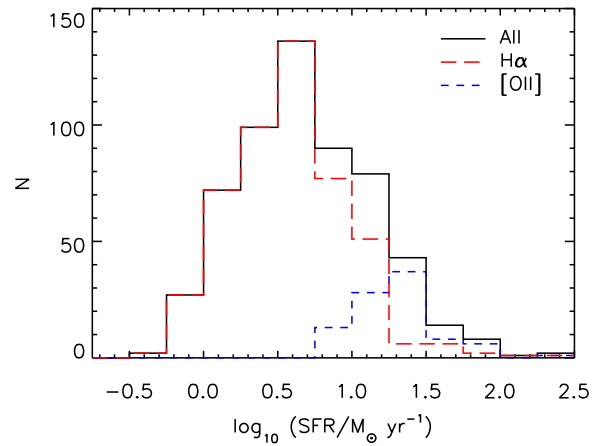}
    \caption{The SFR distribution of the unambiguous-$z$ galaxy sample. The central quasars are not included. The long dashed red line represents the galaxies with H$\alpha$ derived SFR values. The dashed blue line represents the galaxies with [OII] derived SFR values. The black solid line is all SFR values.}
	    \label{fig:sfrhist}
\end{figure}

The star formation rates of {all of} the galaxies {with redshifts} are calculated using the H$\alpha$ and [OII] fluxes, which for the former accounts for [NII] contamination assuming solar metallicity as discussed in \S\ref{sec:spec}. We use the relationships between nebula line flux and star formation from \cite{kennicutt1998} assuming a correction to a \cite{chabrier2003} stellar initial mass function (IMF). A constant extinction of $A_V=1.0$ is also assumed, which is appropriate to $z=1.5$ star-forming galaxies in a similar mass range \citep{sobral2012}. This corresponds to $A_{\rm [OII]}=1.54$ and $A_{\rm H\alpha}=0.818$ using the extinction curve of \cite{cardelli1989}. Fig. \ref{fig:sfrhist} displays the SFR distribution of the QSAGE galaxies. The median H$\alpha$ and [OII] SFRs for the sample are $3.7\pm 0.5$\,M$_\odot$\,yr$^{-1}$ and $19.9\pm 4.9$\,M$_\odot$\,yr$^{-1}$ respectively. We note that due to the wavelength range available, no galaxies have a detection of both H$\alpha$ and [OII] (see Fig. \ref{fig:zhistall}) and so this apparent discrepancy just reflects the fact that the $z>1.9$ sample detected in [OII] has a higher SFR due to the flux limited nature of the observations. It will be possible to refine the SFR with the $A_V$ and metallicity values calculated via SED fitting in Bielby et al. in prep. That paper will also greatly increase the number of SFR values available by breaking the single-line redshift degeneracy as discussed in \S\ref{sec:spec}.

\subsection{Stellar Masses}

\begin{figure}
		\includegraphics[width=\columnwidth]{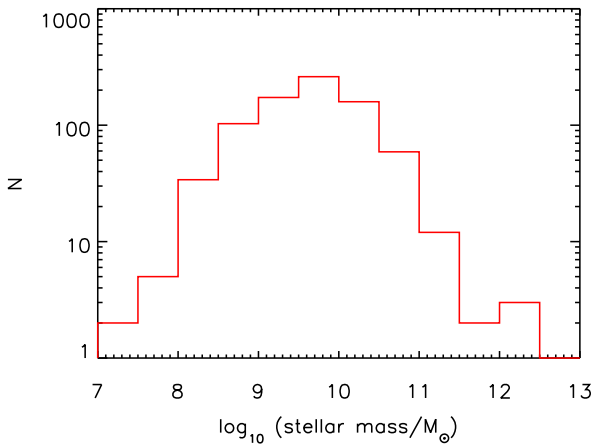}
    \caption{The stellar mass distribution of the QSAGE unambiguous-$z$ galaxy sample. The central quasars are removed from this plot. }
	    \label{fig:masshist}
\end{figure}

The stellar masses {of all of the galaxies with redshifts} are approximately calibrated using the linear relationship between full SED fit stellar mass and F160W magnitude that we derive in bins of 0.1 in redshift from the Cosmic Assembly Near-infrared Deep Extragalactic Legacy Survey (CANDELS) catalogues of \cite{barro2019}. An example of this relation at $z=1.5$ (the approximate median redshift of the QSAGE quasars) is shown in Fig. \ref{fig:masscal}. This calibration is appropriate to make for this paper as stellar mass only features in \S\ref{sec:dep}. The fit parameters we calculate as a function of redshift are given in Table \ref{tab:massfit}. The stellar mass distribution of the QSAGE galaxies based on this approximation is displayed in Fig. \ref{fig:masshist}. The median mass of the sample is $\log (M/M_\odot)=9.68\pm0.02$. An improved set of stellar mass values will be calculated via full SED fitting in Bielby et al., in prep. That paper will also greatly increase the number of mass values available by breaking the single-line redshift degeneracy.

\section{Analysis and Results}
\label{sec:ana}

\subsection{Quasar environments}
\label{sec:env}

\begin{figure*}
		\includegraphics[width=1\columnwidth, trim=30 20 10 35, clip=true]{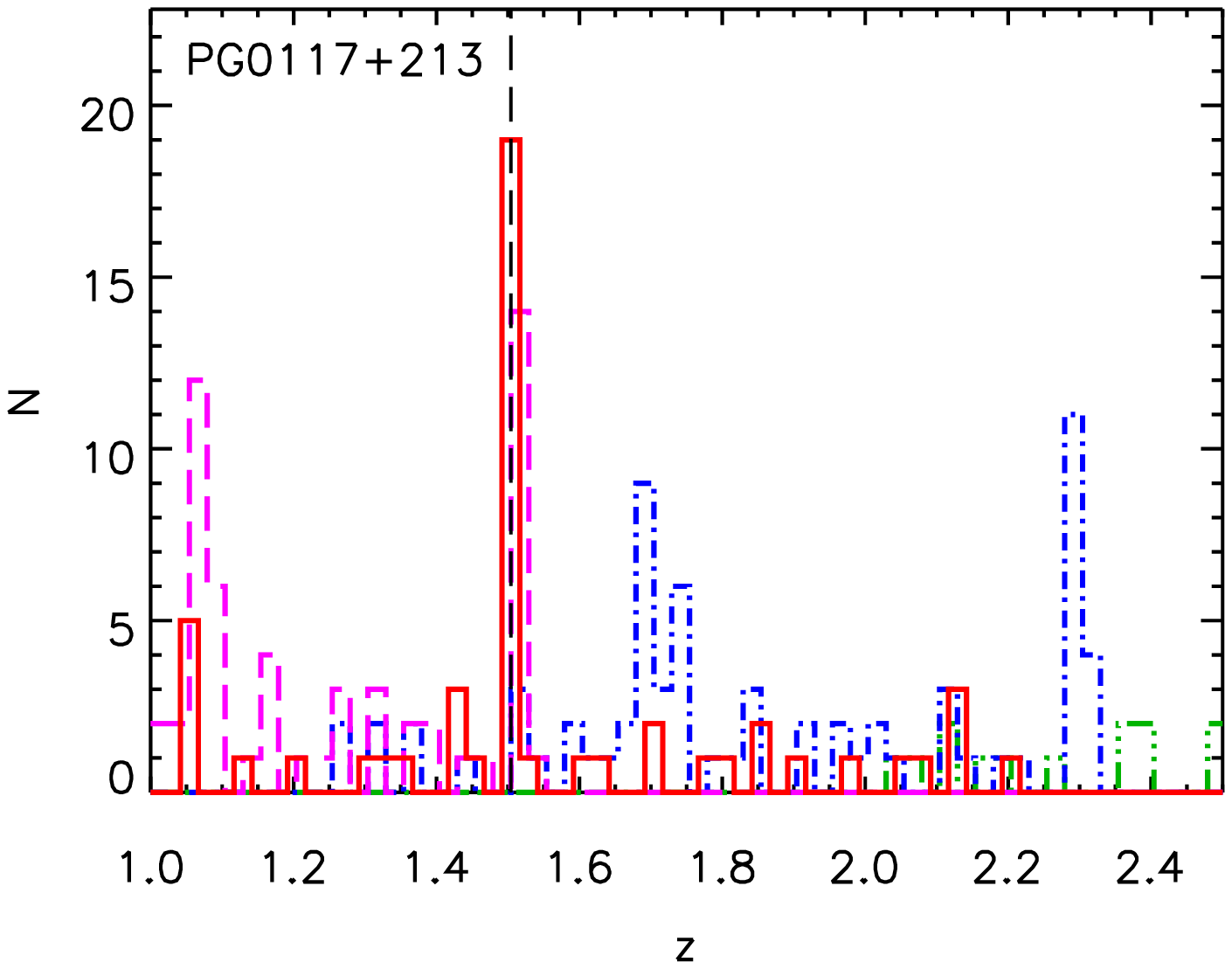}
		\includegraphics[width=1\columnwidth, trim=30 20 10 35, clip=true]{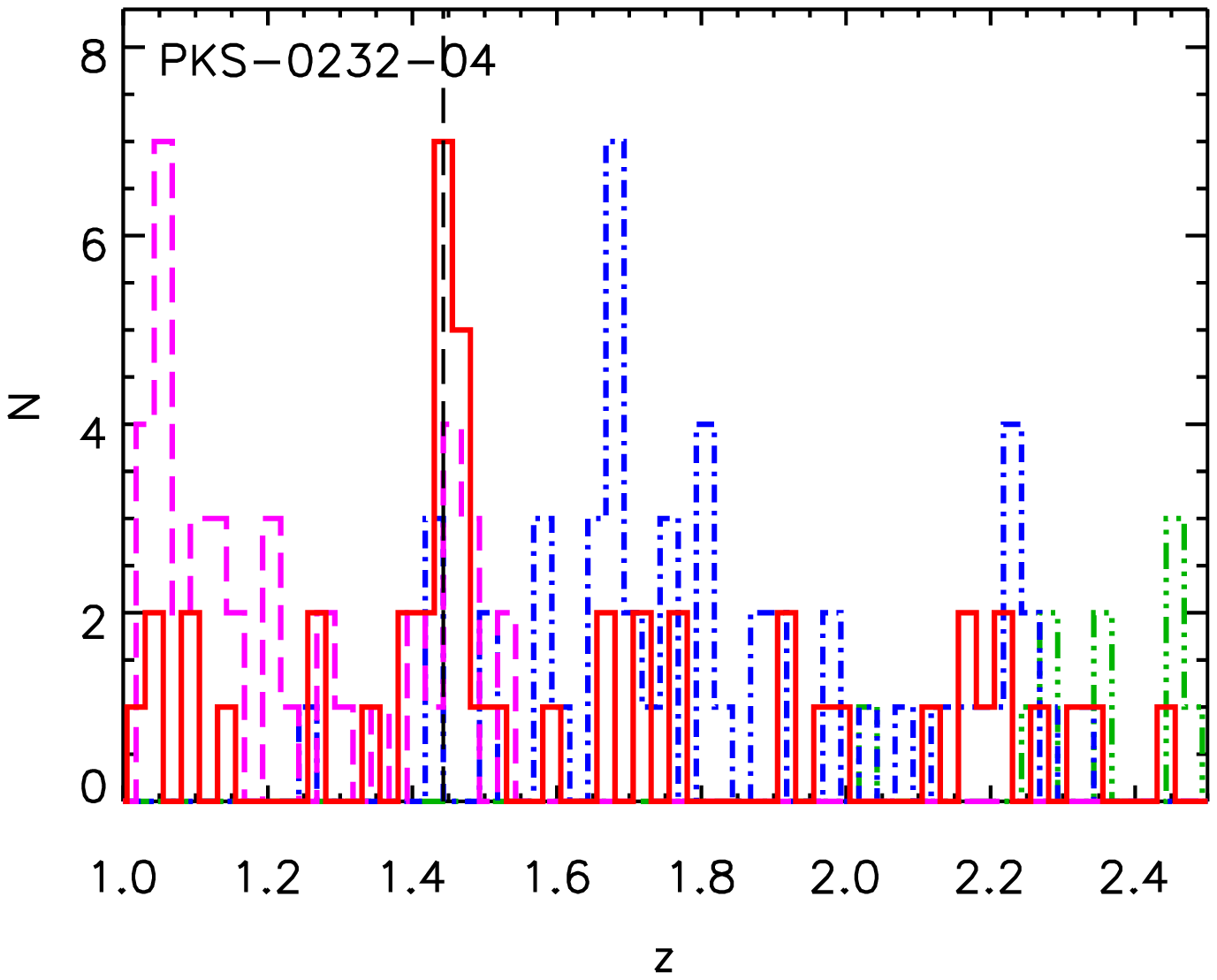}
		\includegraphics[width=1\columnwidth, trim=30 20 10 35, clip=true]{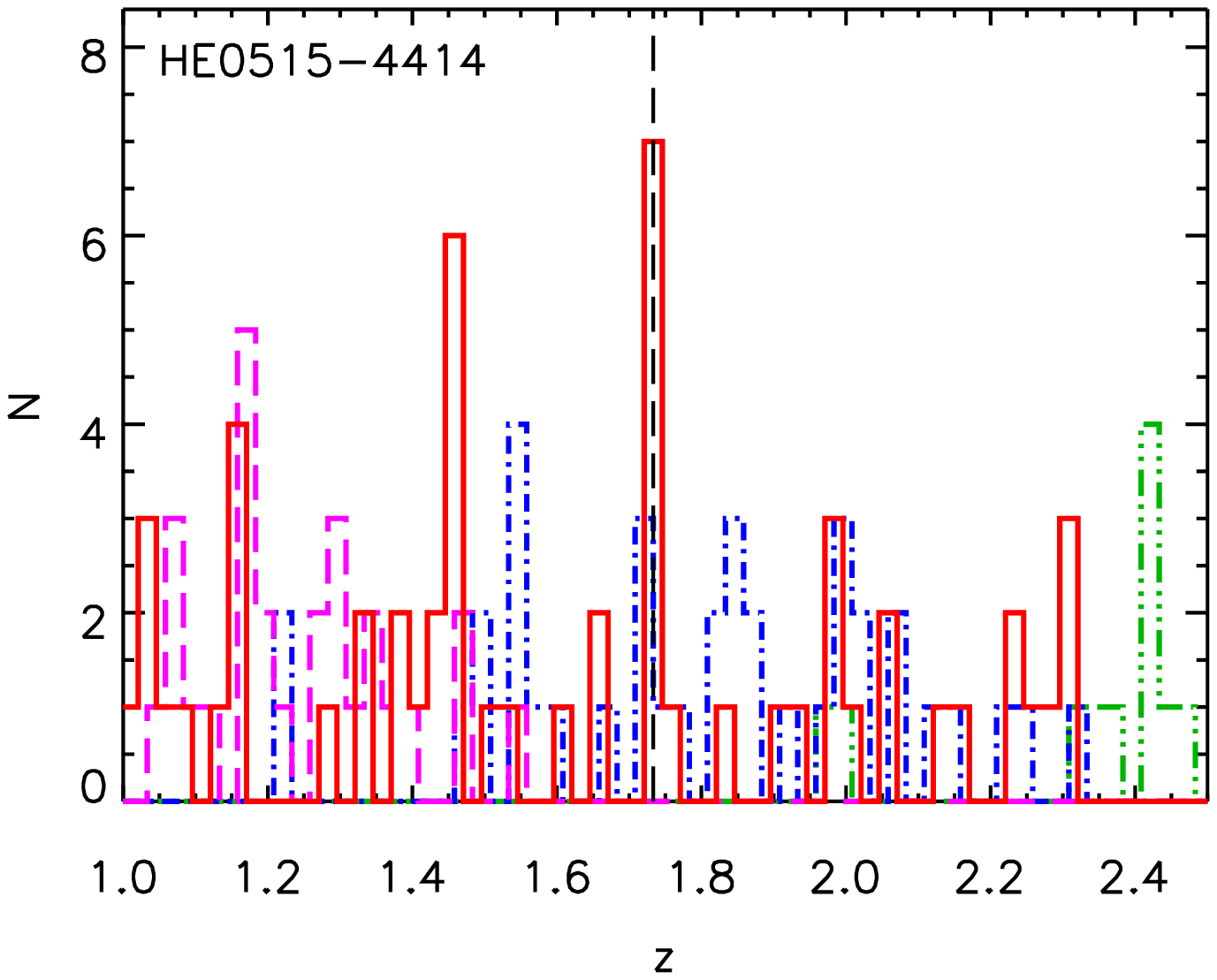}
		\includegraphics[width=1\columnwidth, trim=30 20 10 35, clip=true]{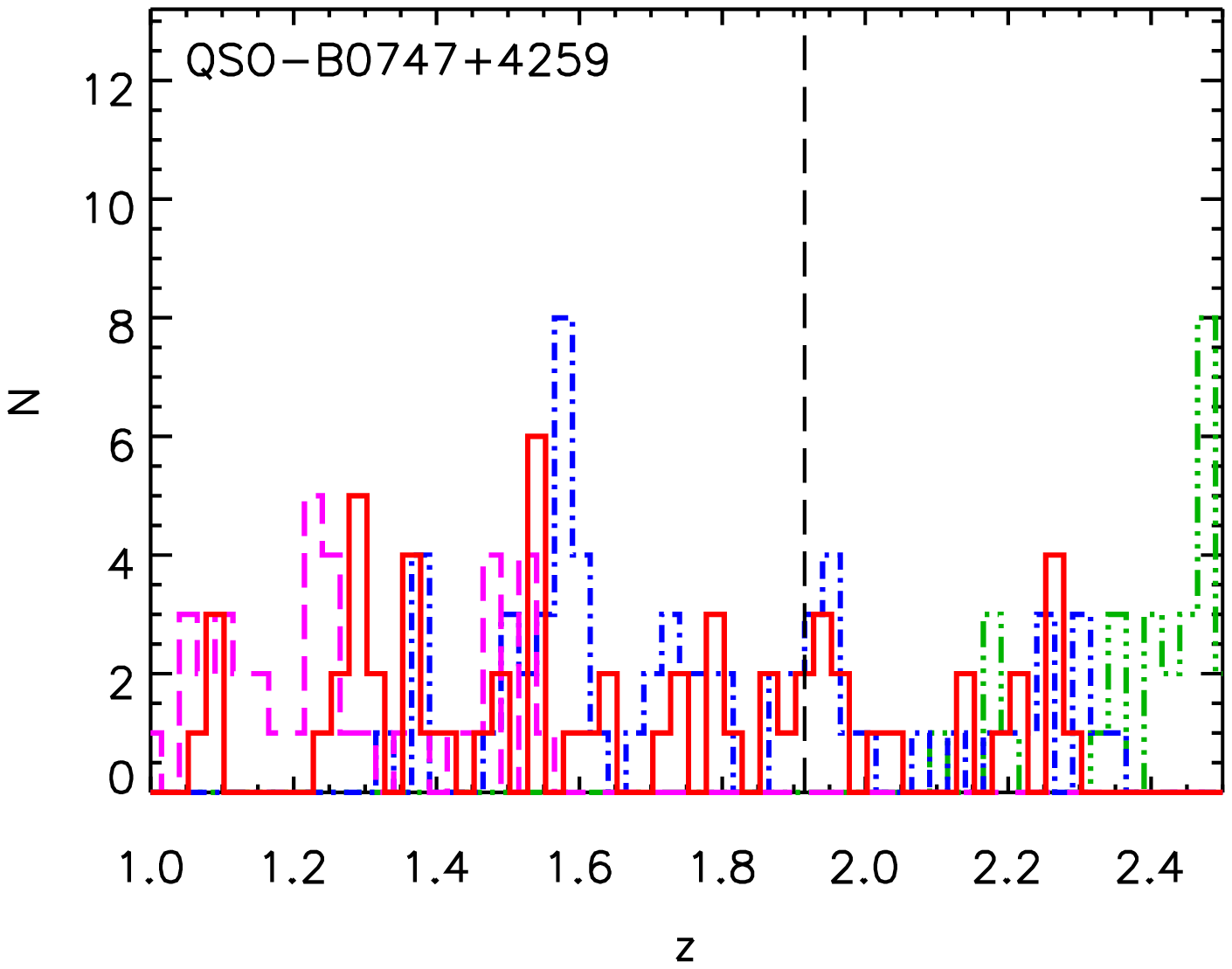}
   		\includegraphics[width=1\columnwidth, trim=30 20 10 35, clip=true]{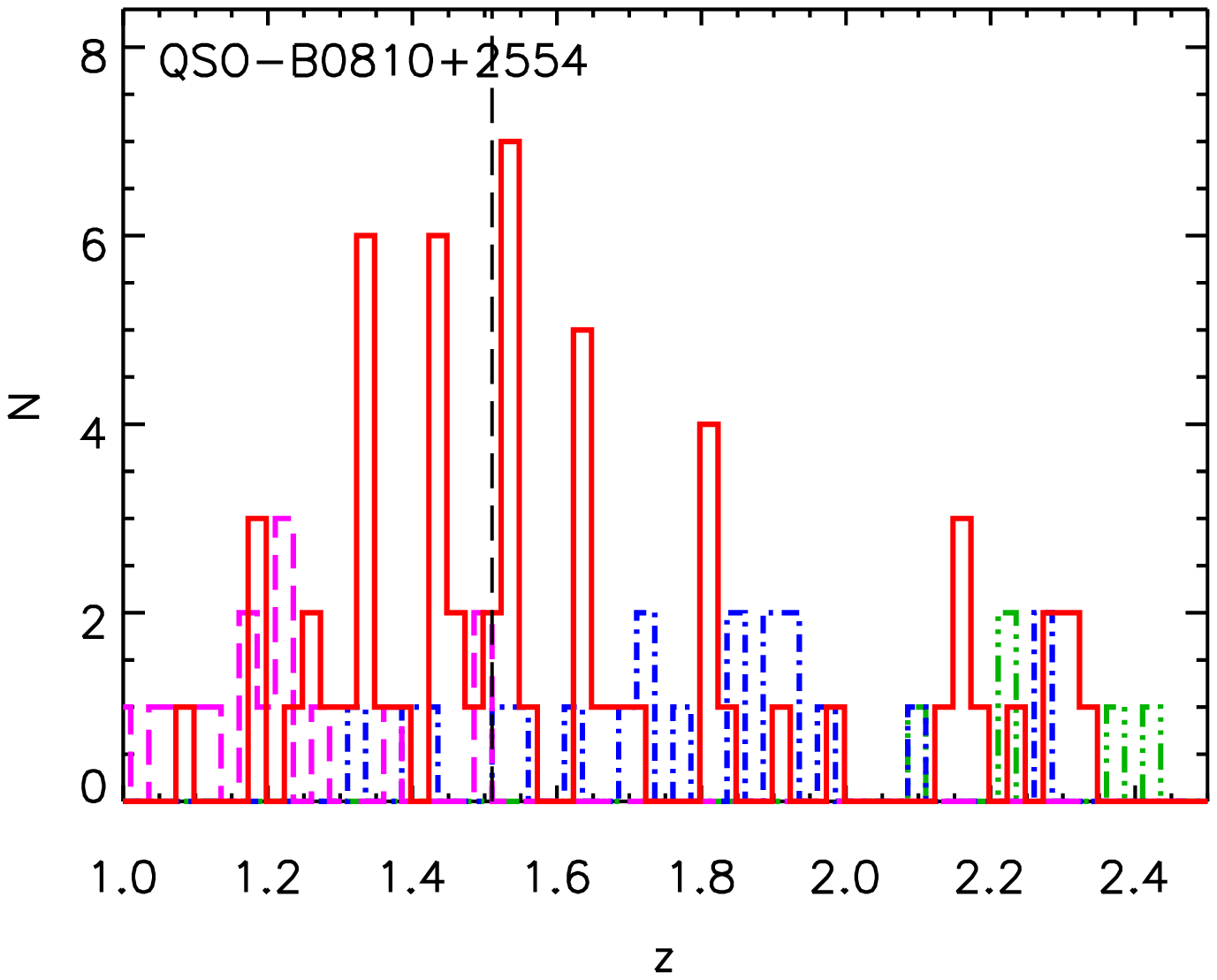}
		\includegraphics[width=1\columnwidth, trim=30 20 10 35, clip=true]{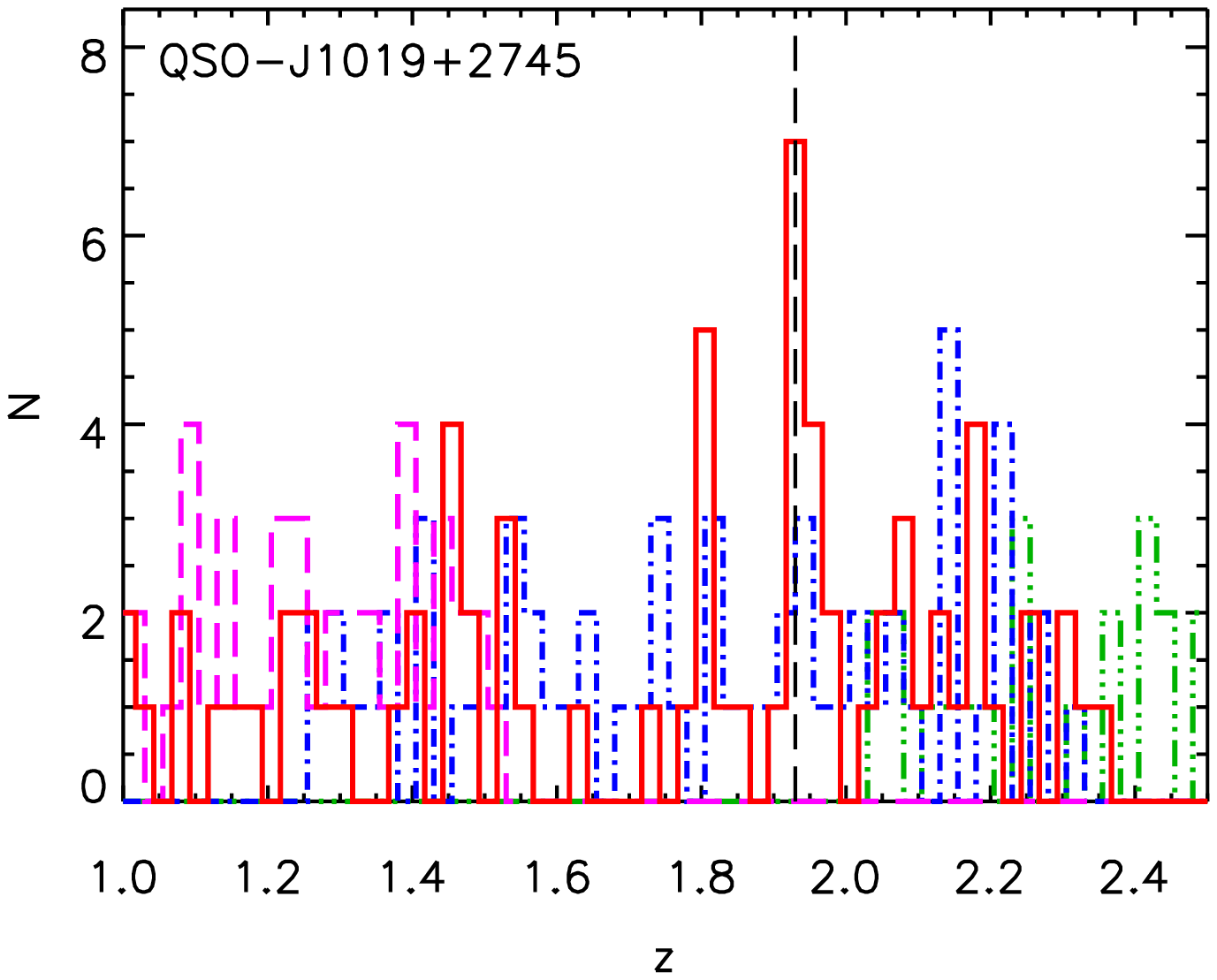}

    \caption{The redshift distribution of the individual QSAGE fields showing the presence of overdensities around 8 of the quasars. The solid red line is formed from the unambiguous-$z$ sample and the black vertical {dashed line is the redshift of the quasar. The dashed magenta, dot-dashed blue and dot-dot-dash green lines} (offset by $\Delta z=0.0125$ for clarity) are the result of assuming that the ambiguous single-line detections are either H$\alpha$, [OIII] or [OII] respectively. This is done to look for potential overdensity members that have been missed as they do not have multiple line detections. One can see that in the case of PG0117+213 there are an additional 17 such galaxies, bringing its total number of members to 36. {See Fig. \ref{fig:zhistind2} for the remaining six fields.}}
	    \label{fig:zhistind}
\end{figure*}

\begin{figure*}
	\includegraphics[width=1\columnwidth, trim=30 20 10 35, clip=true]{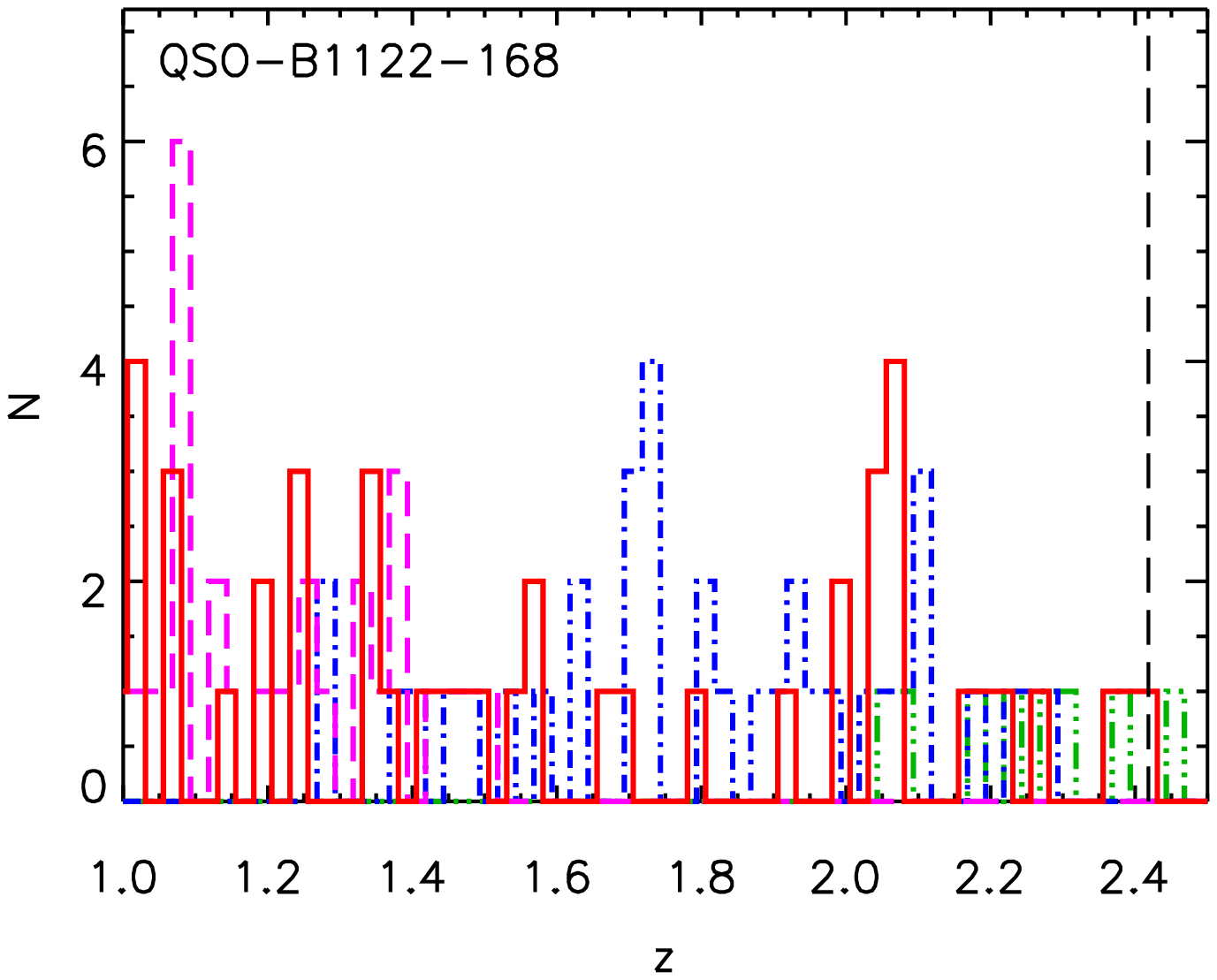}
		\includegraphics[width=1\columnwidth, trim=30 20 10 35, clip=true]{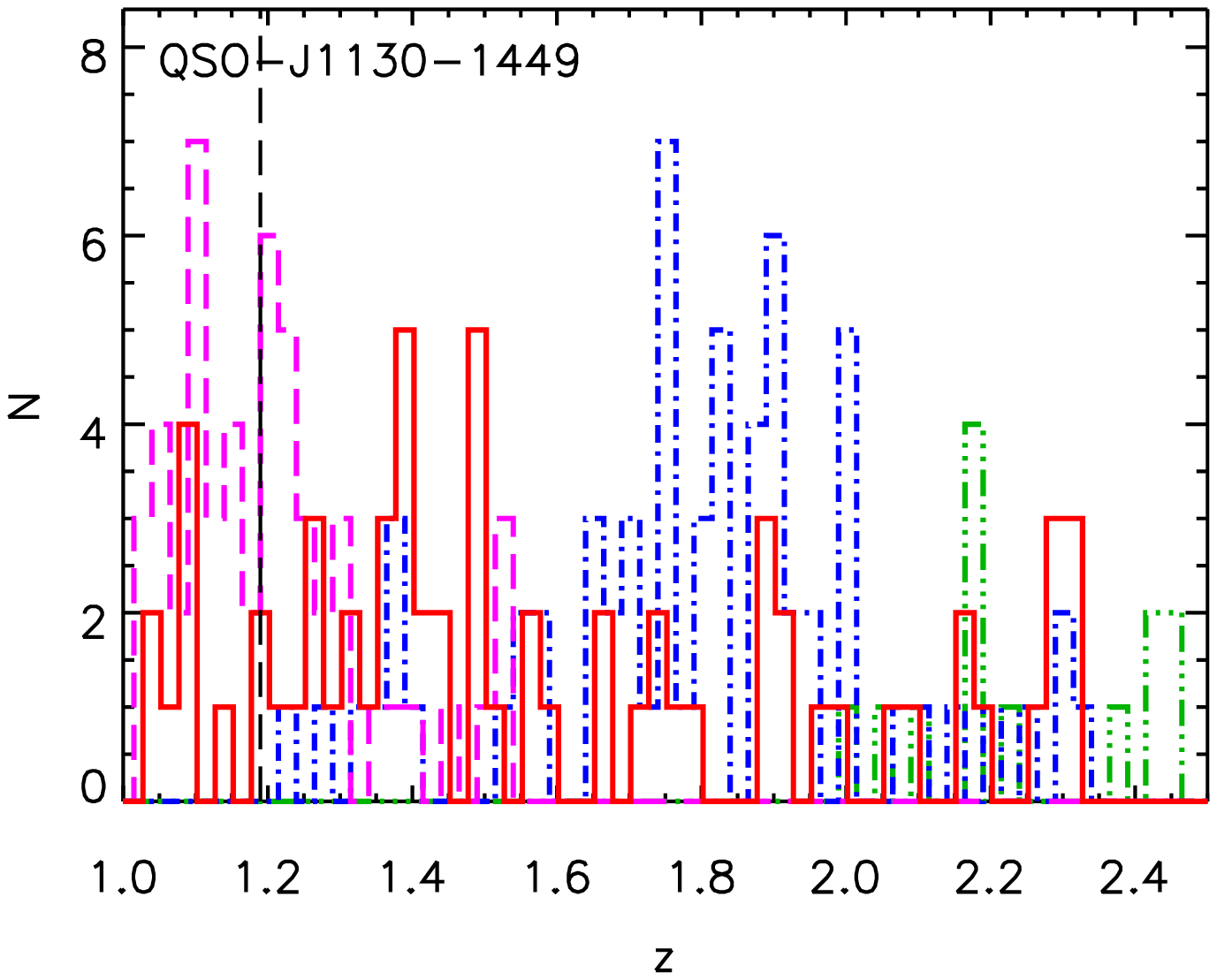}
		\includegraphics[width=1\columnwidth, trim=30 20 10 35, clip=true]{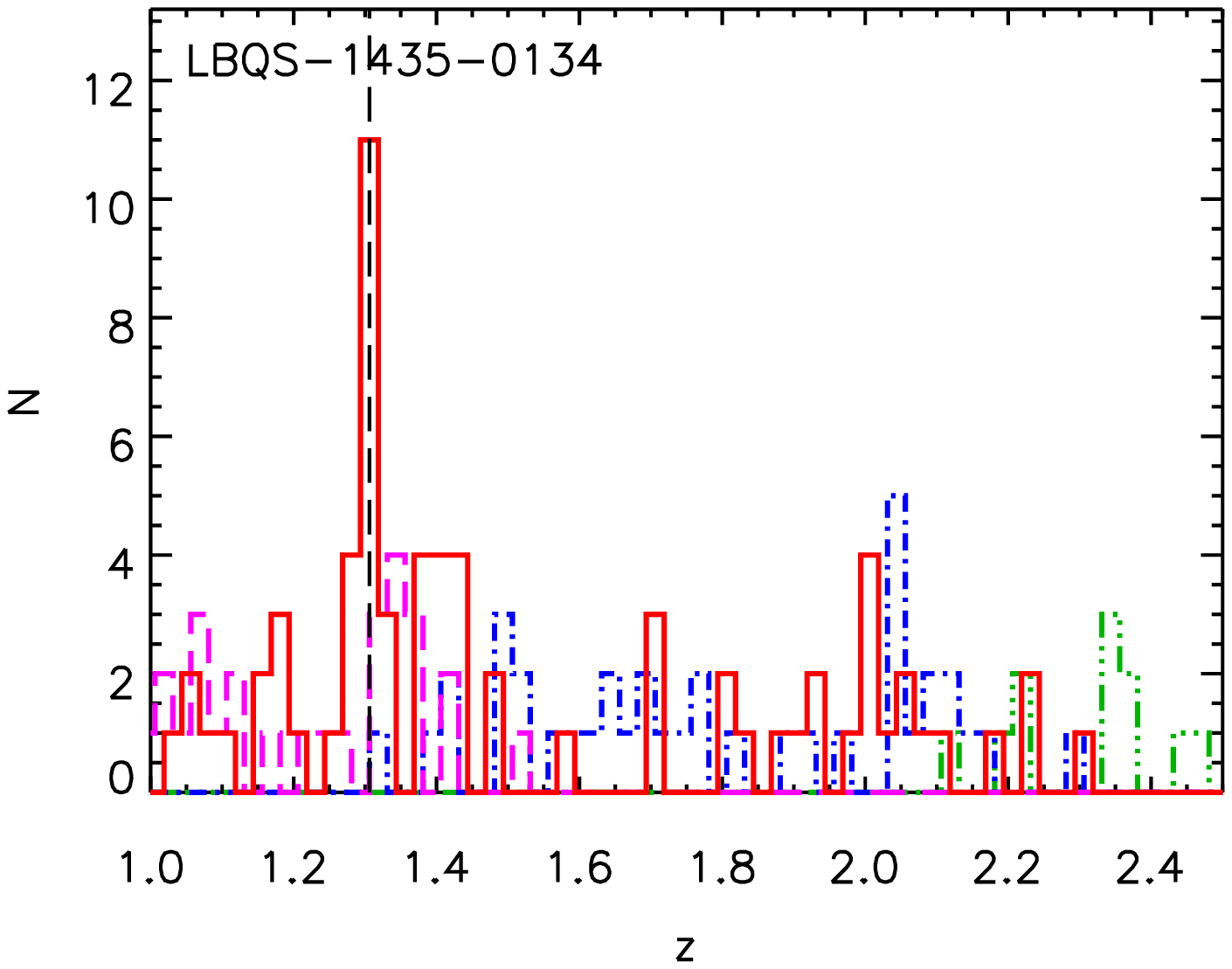}
		\includegraphics[width=1\columnwidth, trim=30 20 10 35, clip=true]{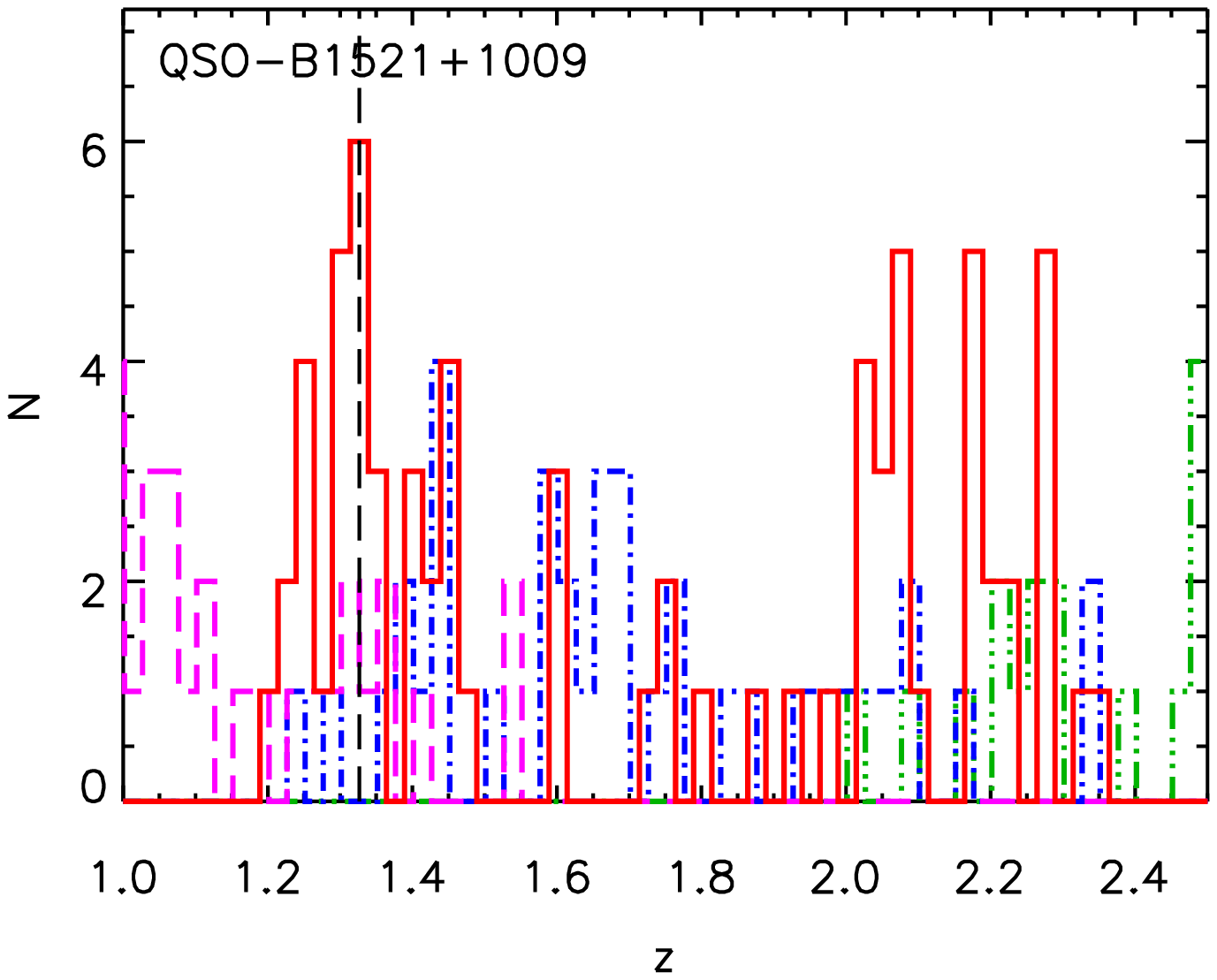}
		\includegraphics[width=1\columnwidth, trim=30 20 10 35, clip=true]{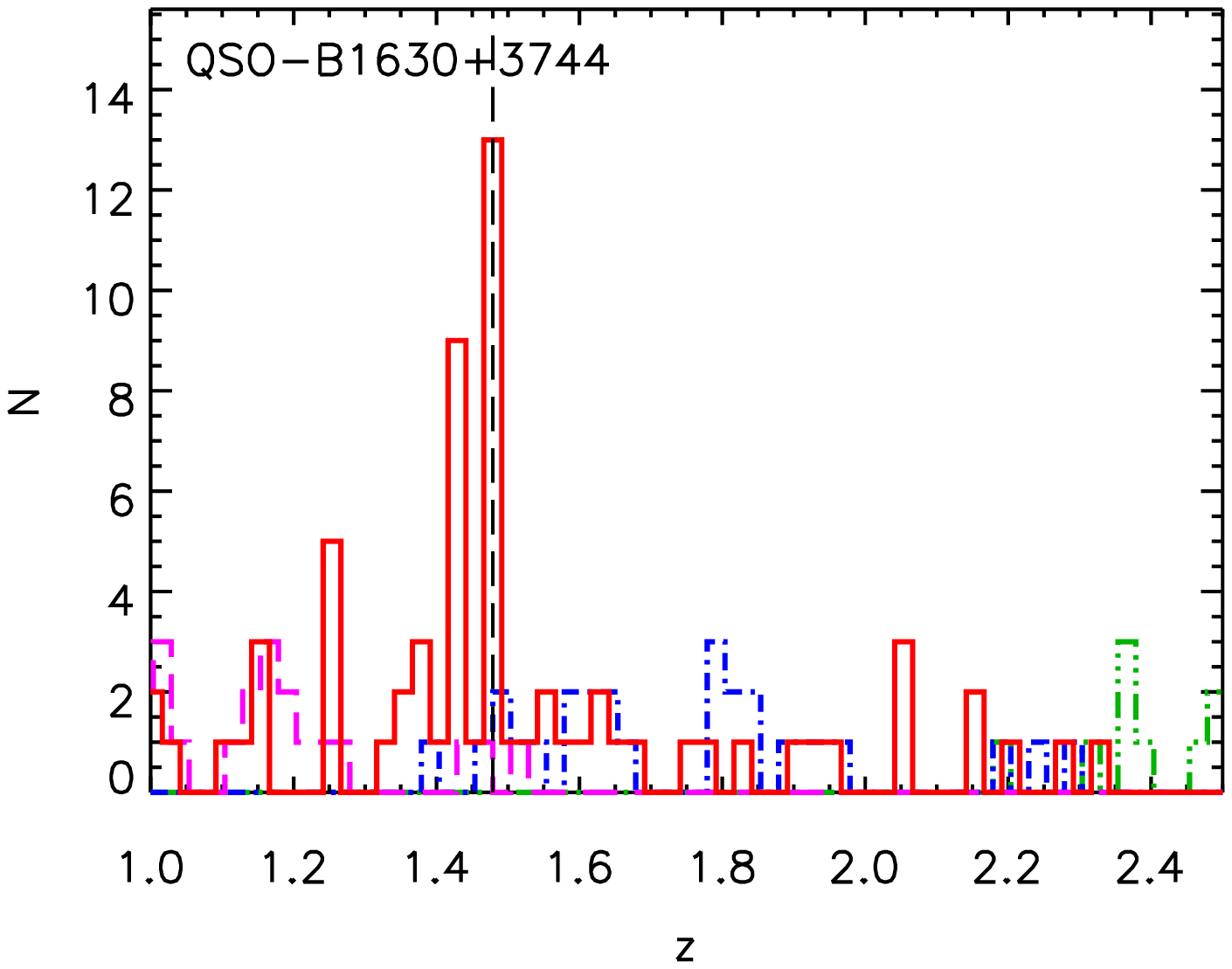}
		\includegraphics[width=1\columnwidth, trim=30 20 10 35, clip=true]{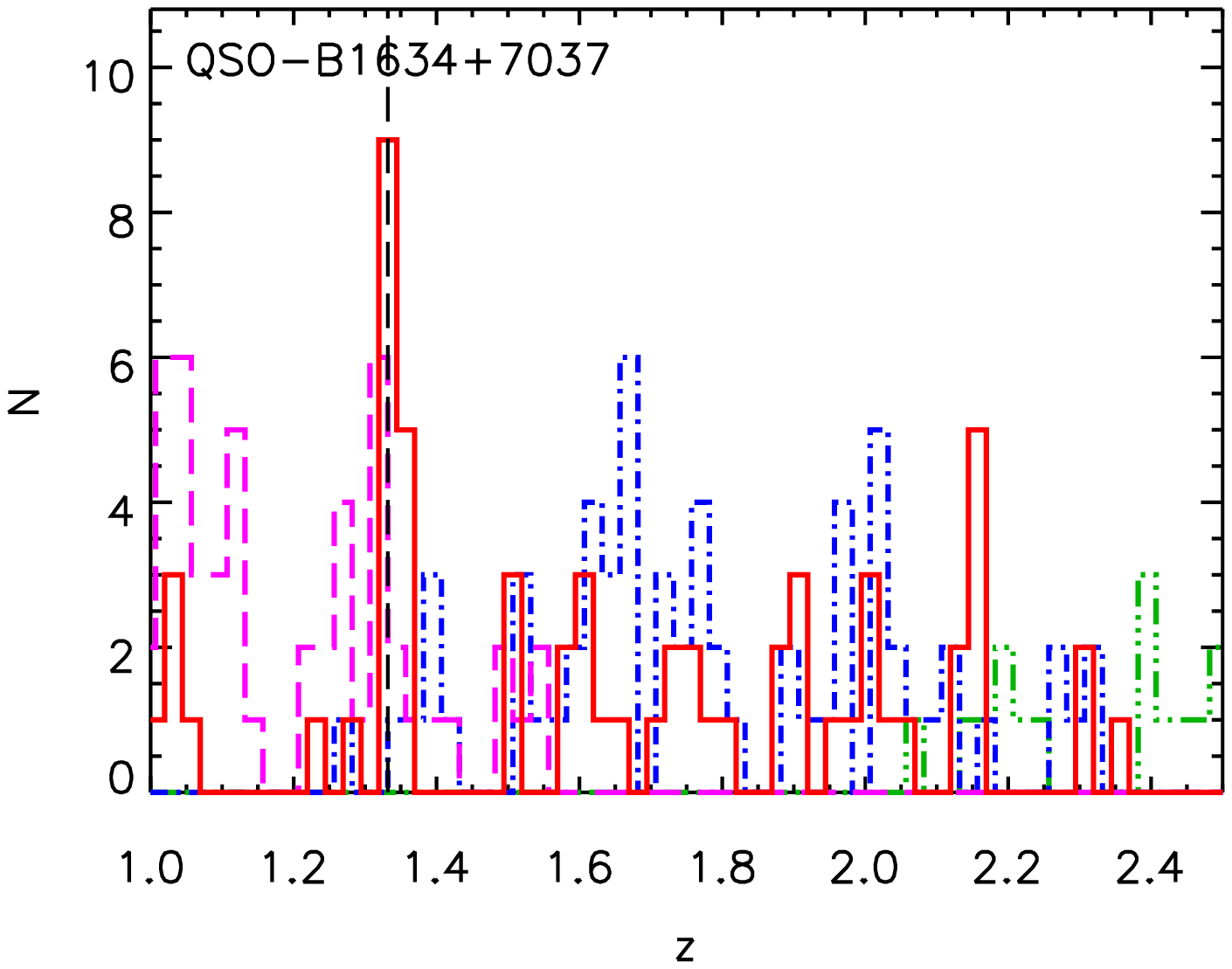}
		\caption{A continuation of Fig. \ref{fig:zhistind}.}
	    \label{fig:zhistind2}
\end{figure*}

We search for the presence of overdensities in the redshift distribution in the vicinity of the central QSAGE target quasars. The advantage of using a grism for this analysis is that we obtain an unbiased sample of line-emitting galaxies as there is no preselection of spectroscopic targets. While this is also true for previous narrow-band surveys (e.g., \citealt{lefevre1996,kurk2004,husband2016}) we are not restricted to only studying quasars at the redshifts in which a narrow-band filter can isolate an emission line. In addition, our unambiguous-$z$ sample is unaffected by interlopers due to alternative spectral lines from galaxies at different redshifts being detected in the narrow-band filter. Our method for finding overdensities is to search for an excess in galaxies in a fixed $\Delta z=0.025$ redshift bin centred on the measured WFC3 grism quasar redshift. The choice of $\Delta z=0.025$ reflects a combination in quadrature of the spectral resolution of the grism (instrumental velocity dispersion $\sigma_{\rm inst}=c/(R\times 2.35)\approx 1000$\,km\,s$^{-1}$) and the typical velocity dispersion of a forming cluster ($\sigma_{\rm clust}=3000$\,km\,s$^{-1}$) and a factor of $(1+z)$ to account for the expansion of the Universe (using the approximate average quasar redshift, $z=1.5$). The individual redshift distributions are presented in Figures \ref{fig:zhistind} and \ref{fig:zhistind2}. 

To assess the environment of the quasars we use the overdensity parameter, $\delta_g$, which is given by

\begin{equation}
   \delta_g=\frac{N_{\rm clust}-N_{\rm bg}}{N_{\rm bg}}
	\label{eq:overdense}
\end{equation}

\noindent where $N_{\rm clust}$ is the number of unambiguous-$z$ galaxies in the $\Delta z=0.025$ bin centred on the redshift of the quasar and $N_{\rm bg}$ is the average background i.e. the expected number of galaxies in the absence of any significant structure. The calculation of this background value, $N_{\rm bg}$, is therefore key to understanding the significance of the galaxy overdensity. 

The overdensity parameter is often used when the presence of an excess of galaxies is being assessed with narrow-band imaging (e.g. \citealt{husband2016}). In that case, the number of narrow-band selected galaxies at the redshift of the quasar/radio galaxy is compared with the number of galaxies selected in the same narrow-band (to the same magnitude limit) when averaged over a large area or from `blank' fields. In our case, we are searching for potential overdensities at 12 different redshifts in fields in which galaxies in a similar redshift range are detected and selected in the same way. This can be used to our advantage as it means we can generate an average background in redshift space using adjacent redshift bins to the quasar itself in combination with those from the remaining 11 quasar fields.  

\begin{figure*}
		\includegraphics[width=0.64\columnwidth, viewport=417 51 1388 1056, clip=true]{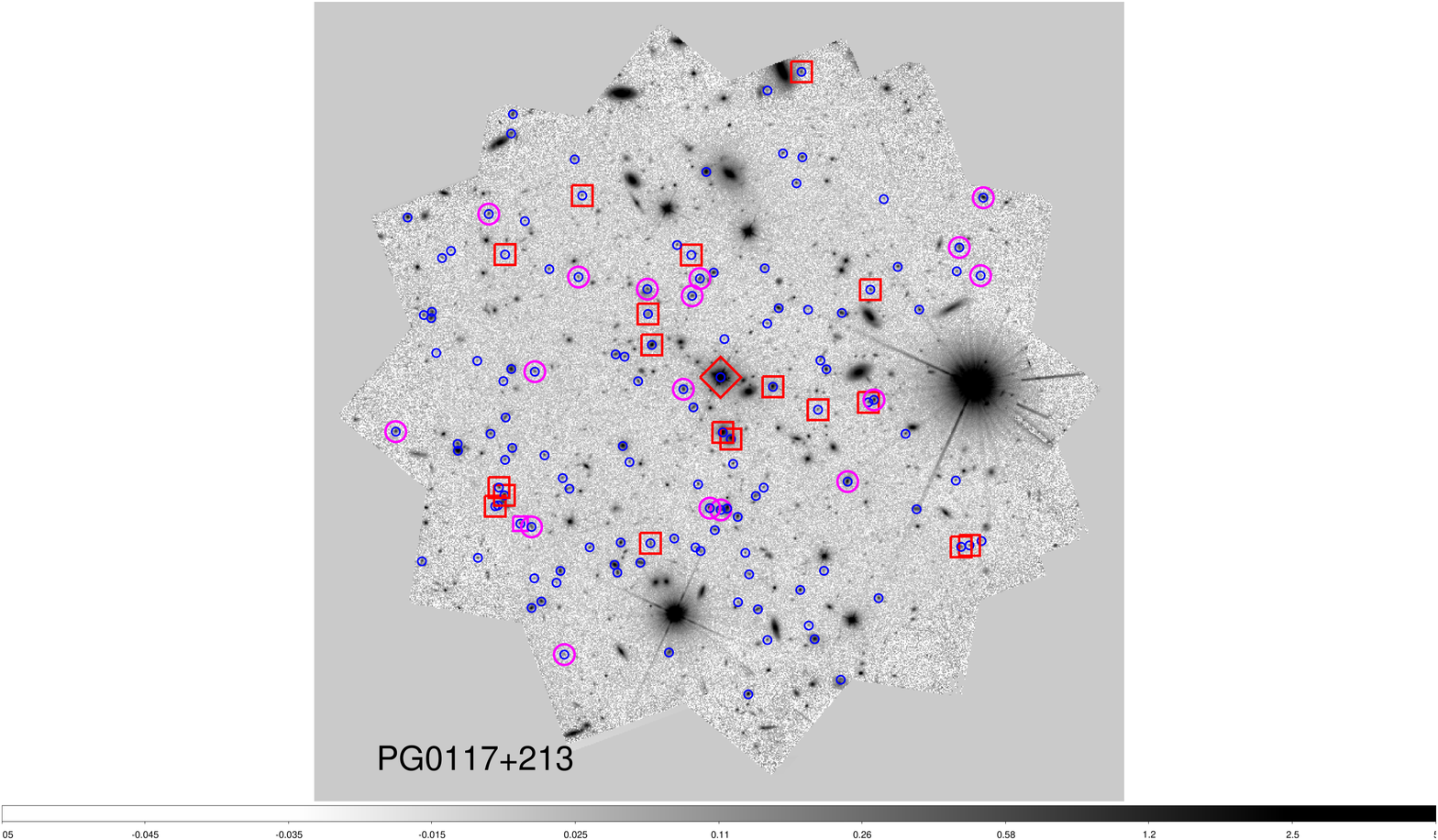}
		\includegraphics[width=0.64\columnwidth, viewport=417 51 1388 1056, clip=true]{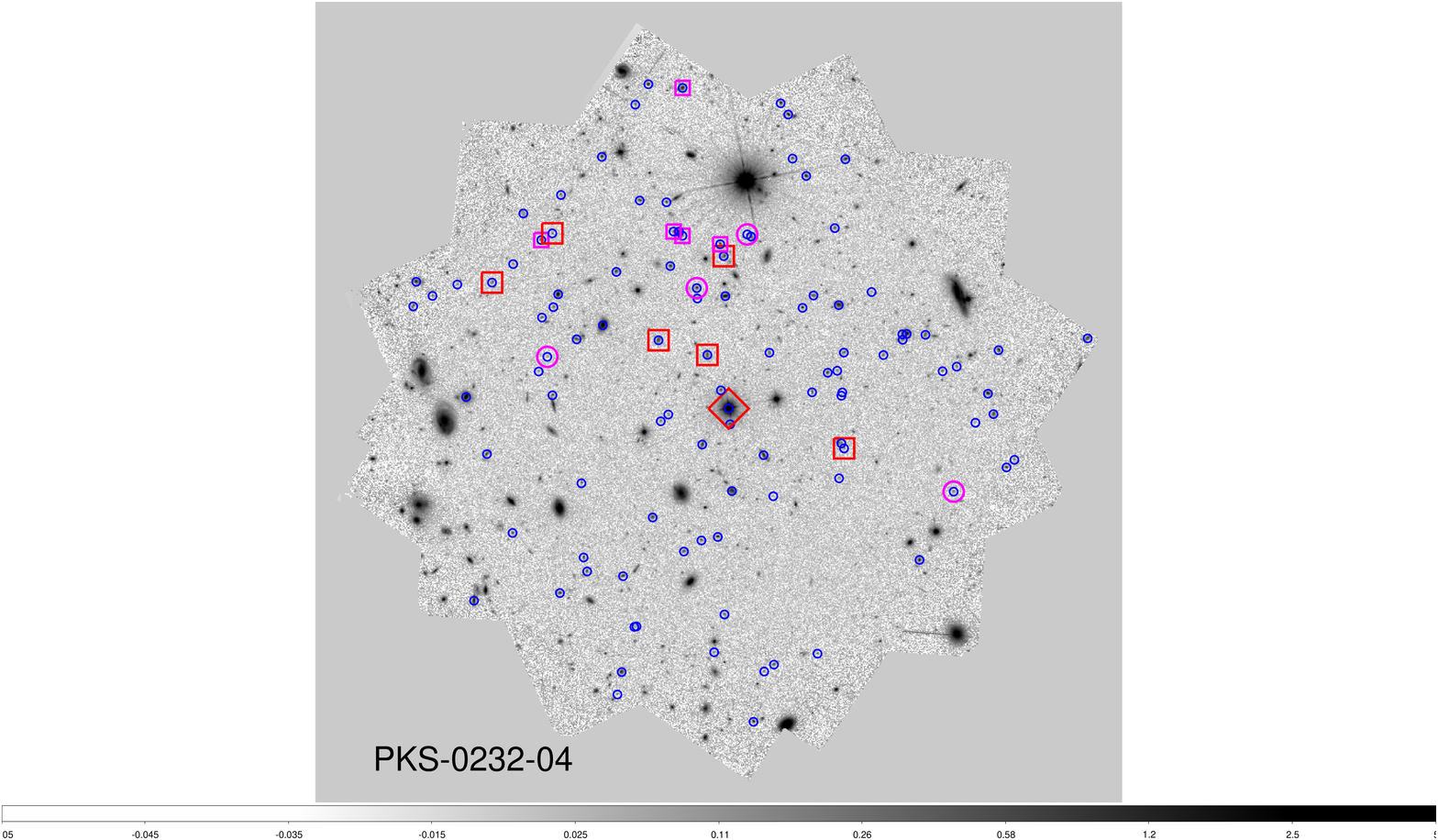}
		\includegraphics[width=0.64\columnwidth, viewport=417 51 1388 1056, clip=true]{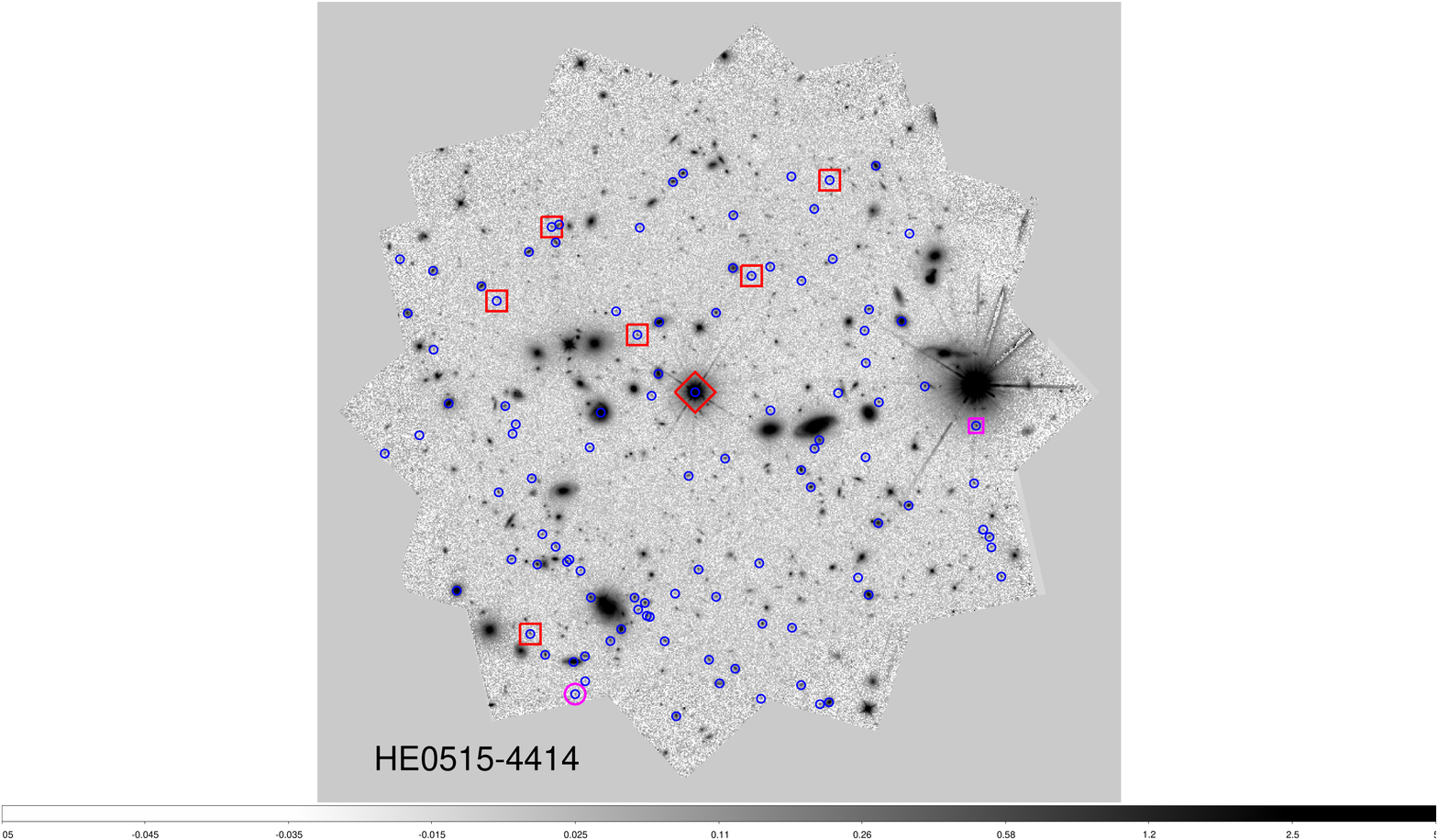}
		\includegraphics[width=0.64\columnwidth, viewport=390 51 1361 1056, clip=true]{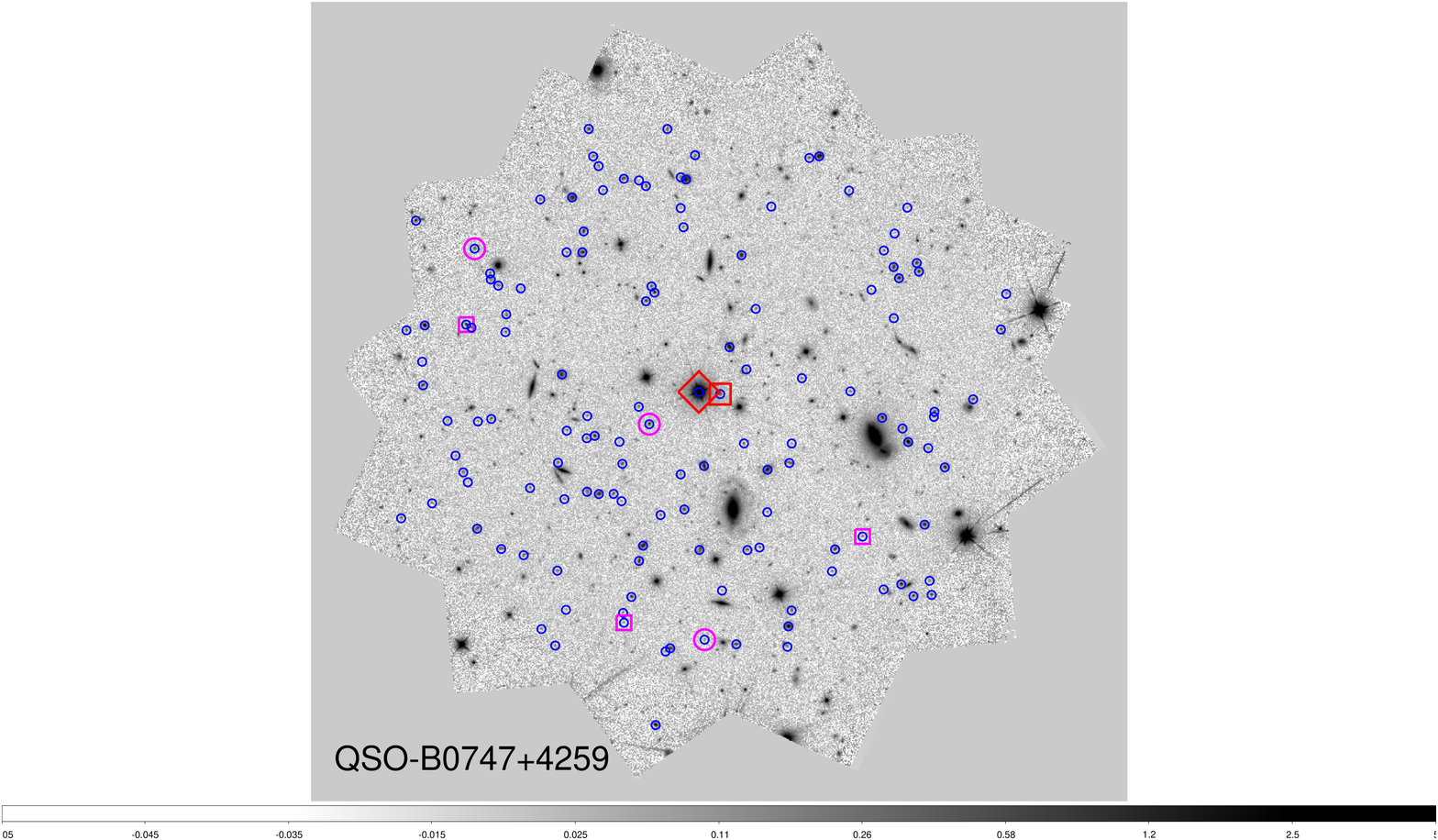}
		\includegraphics[width=0.64\columnwidth, viewport=404 51 1375 1056, clip=true]{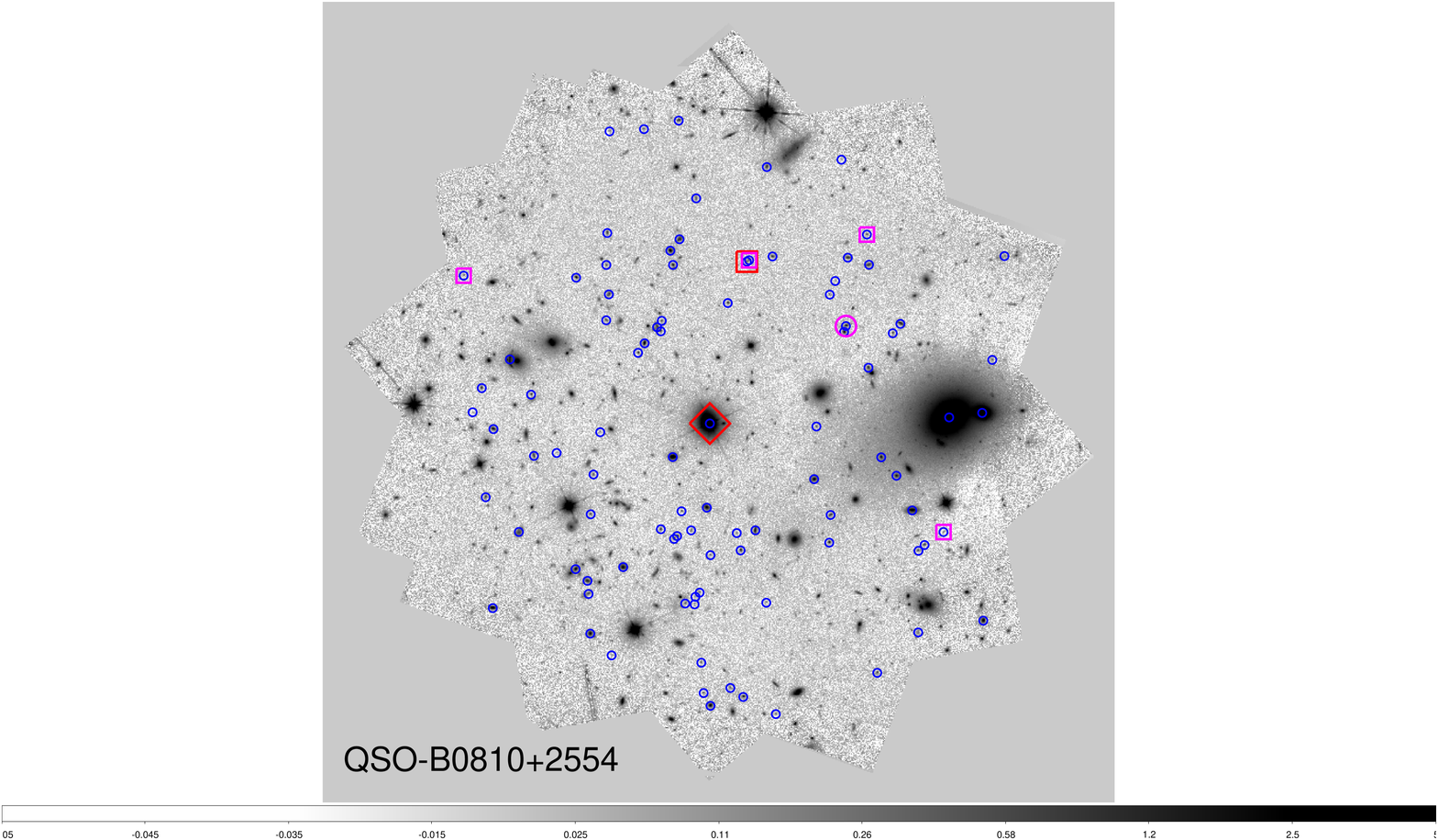}
		\includegraphics[width=0.64\columnwidth, viewport=417 51 1388 1056, clip=true]{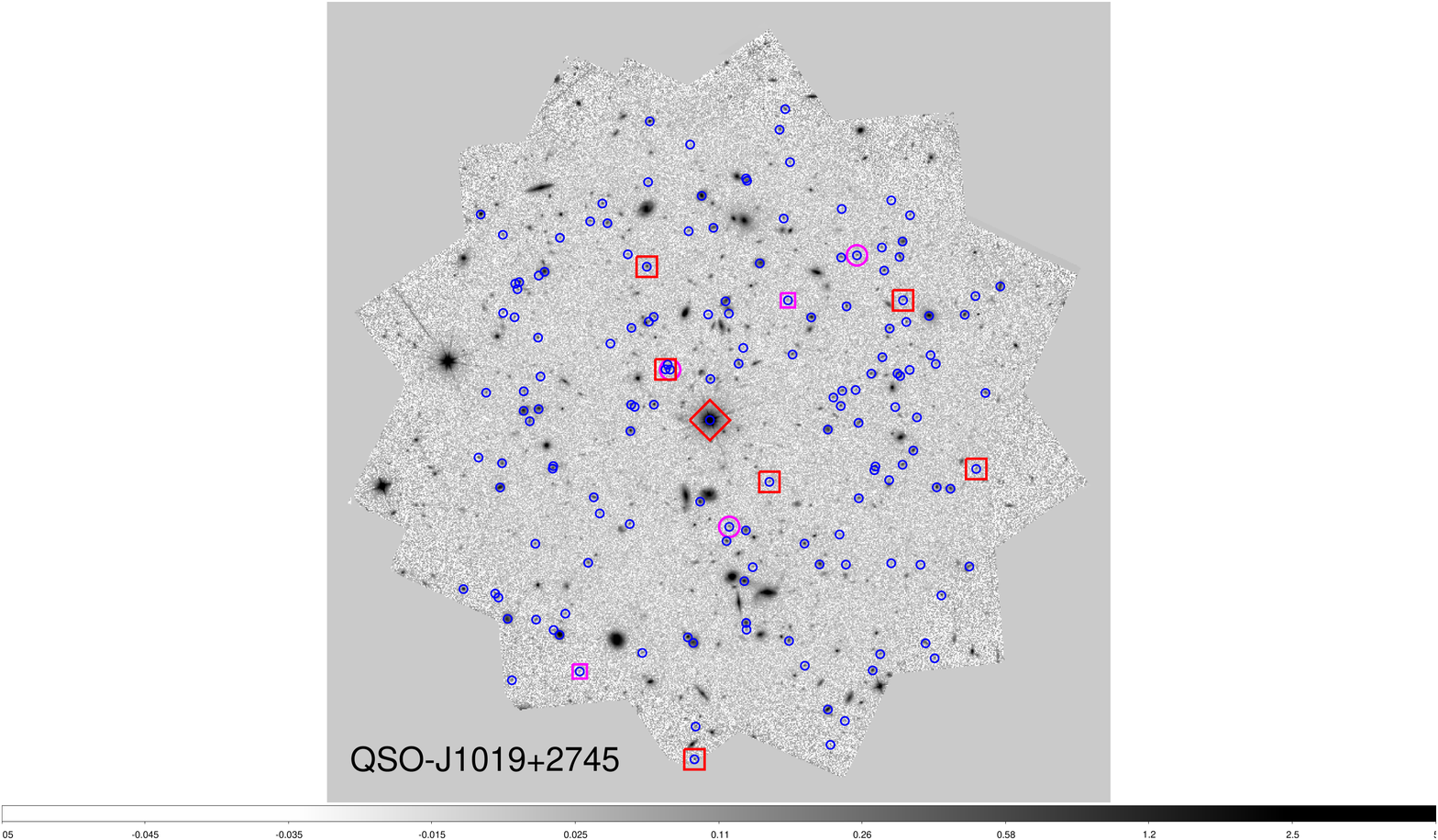}
		\includegraphics[width=0.64\columnwidth, viewport=417 51 1388 1056, clip=true]{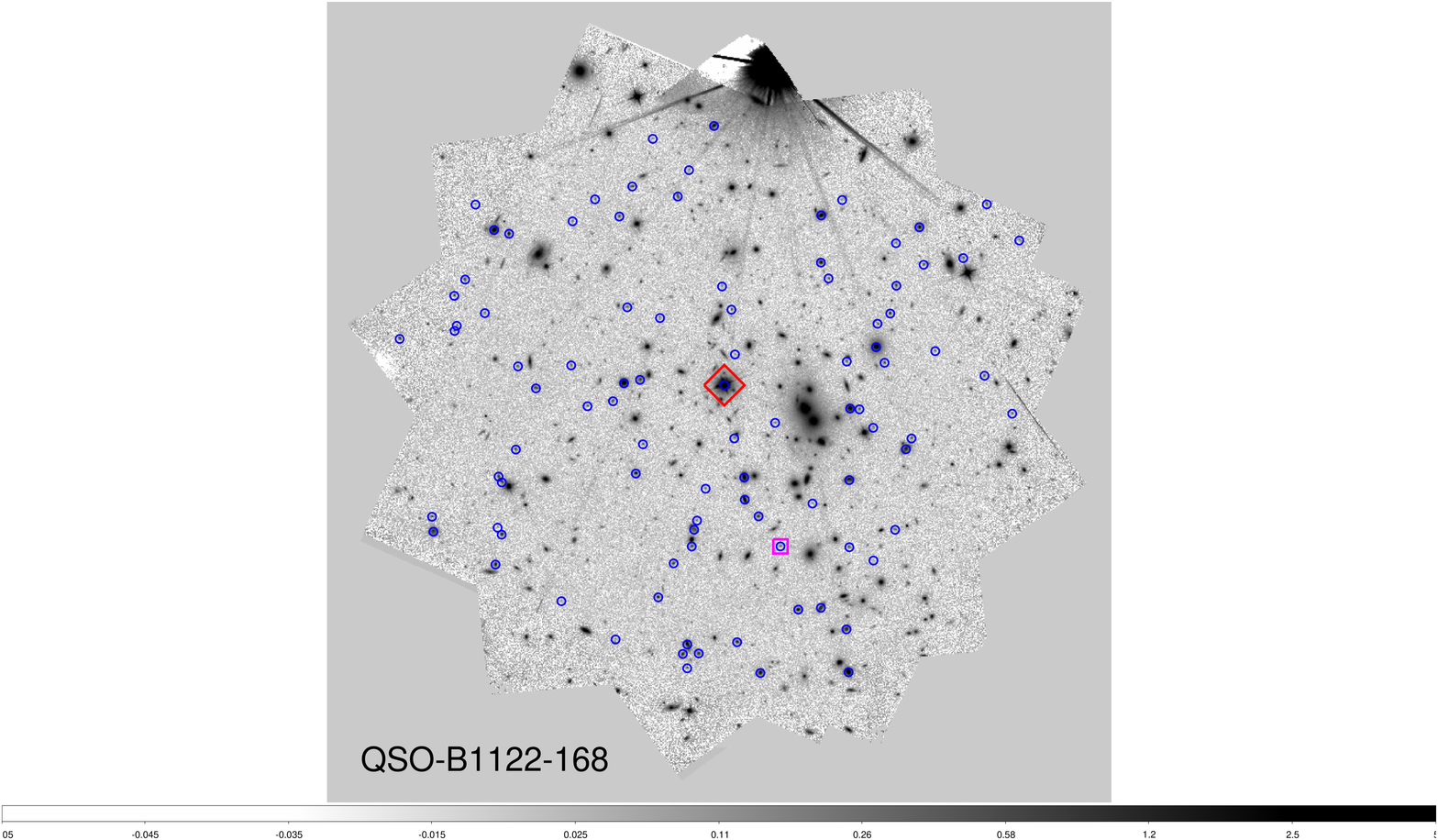}
		\includegraphics[width=0.64\columnwidth, viewport=417 51 1388 1056, clip=true]{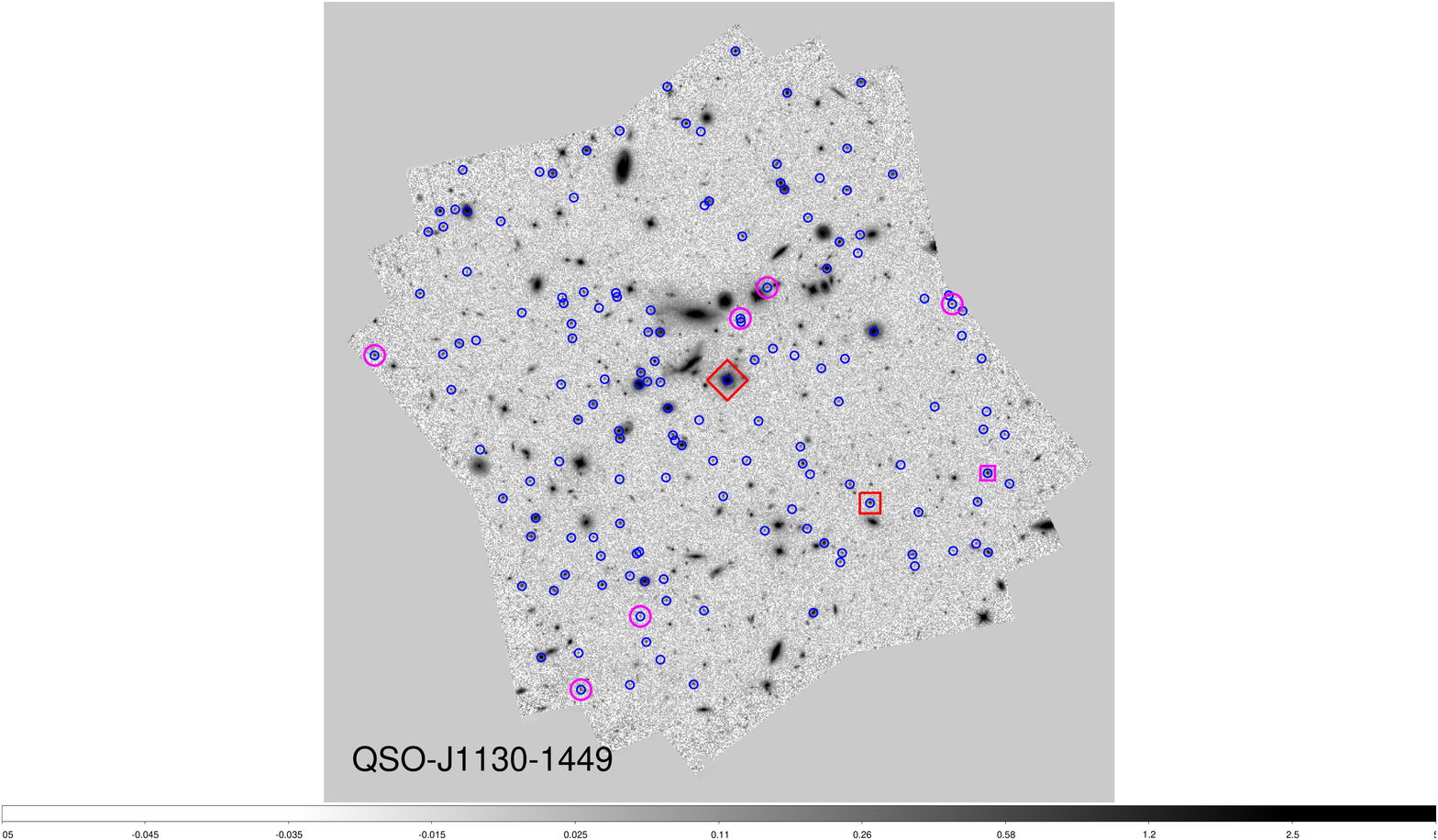}
		\includegraphics[width=0.64\columnwidth, viewport=417 51 1388 1056, clip=true]{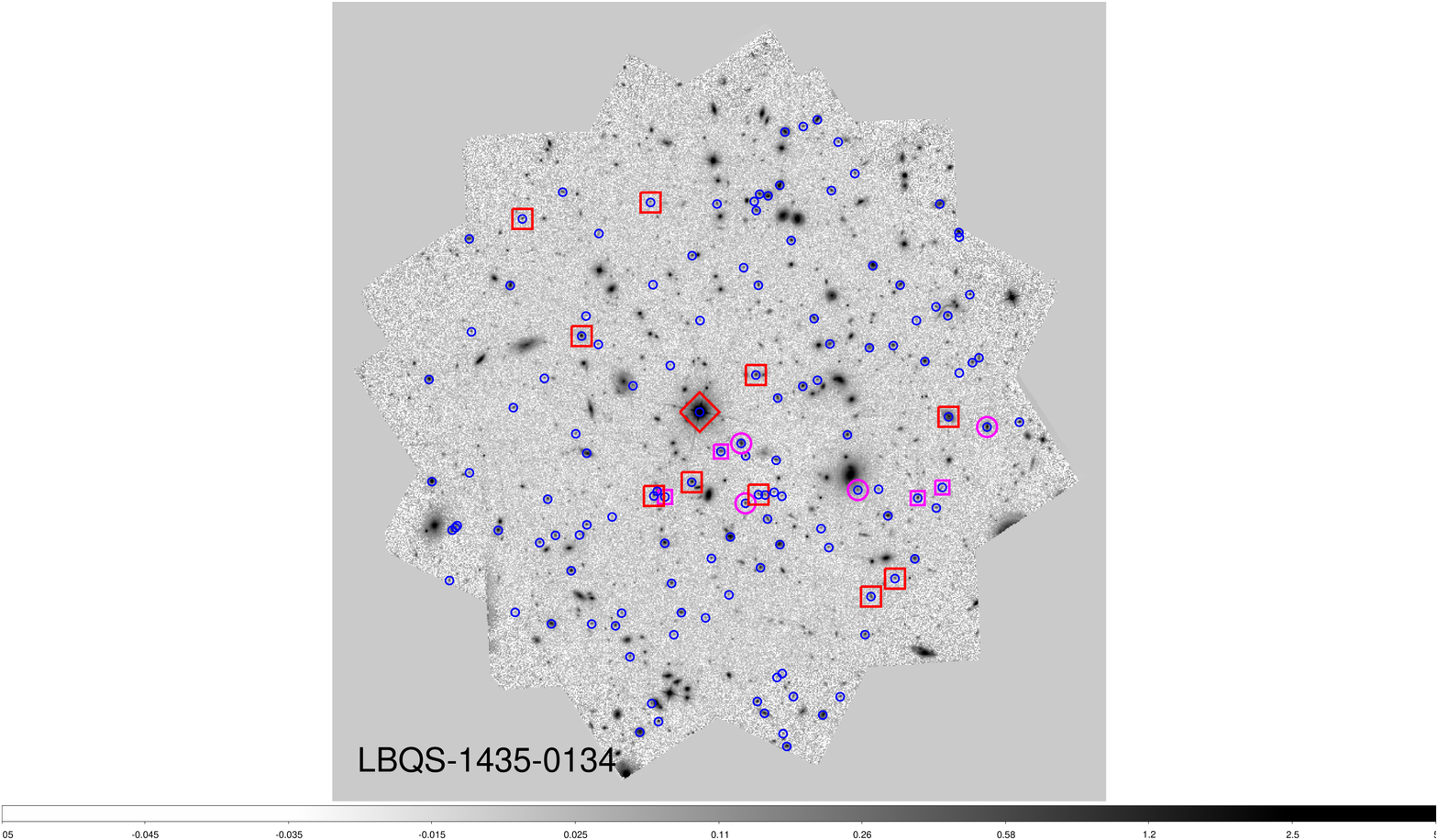}
		\includegraphics[width=0.64\columnwidth, viewport=398 51 1368 1056, clip=true]{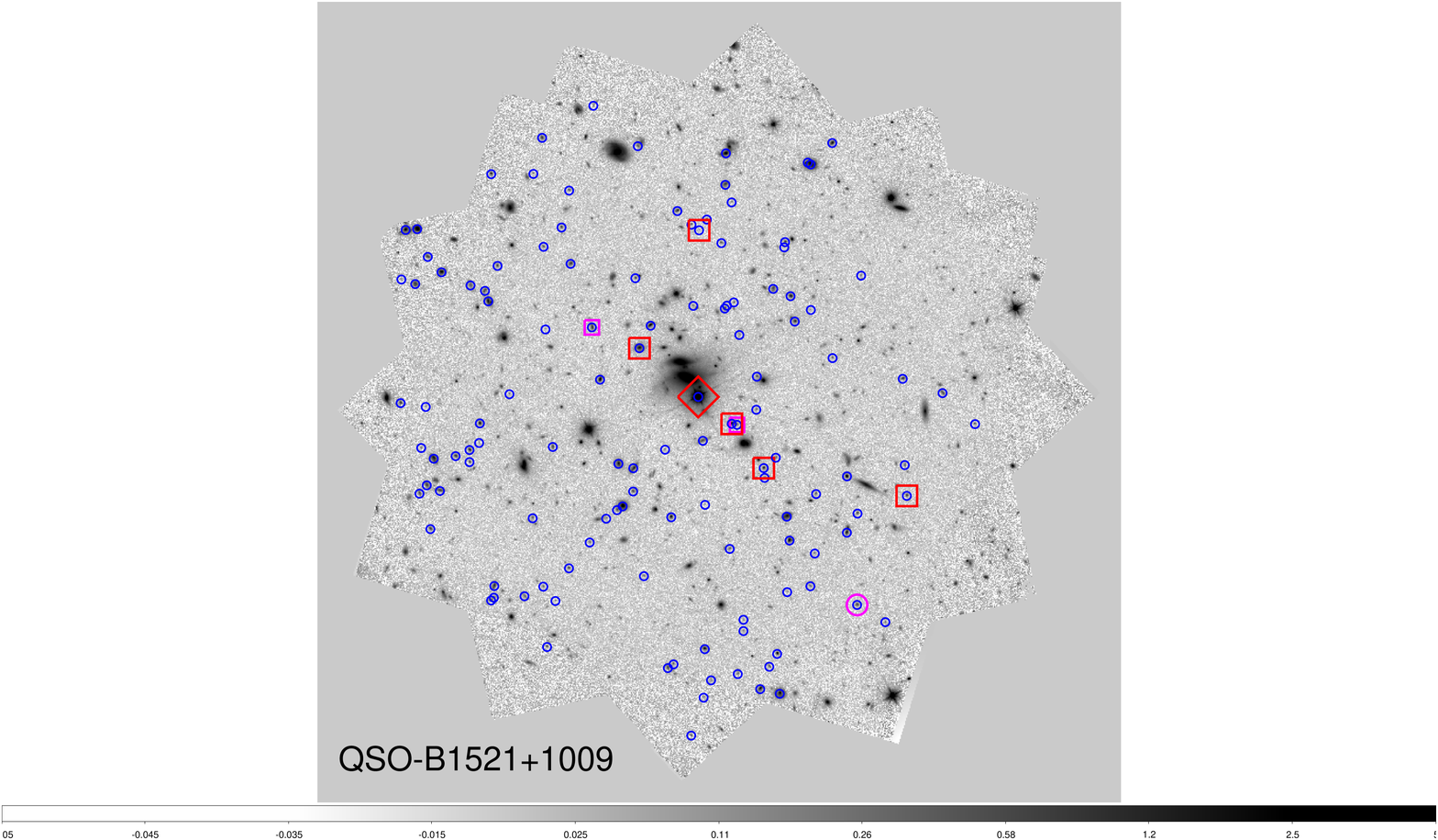}
		\includegraphics[width=0.64\columnwidth, viewport=398 51 1368 1056, clip=true]{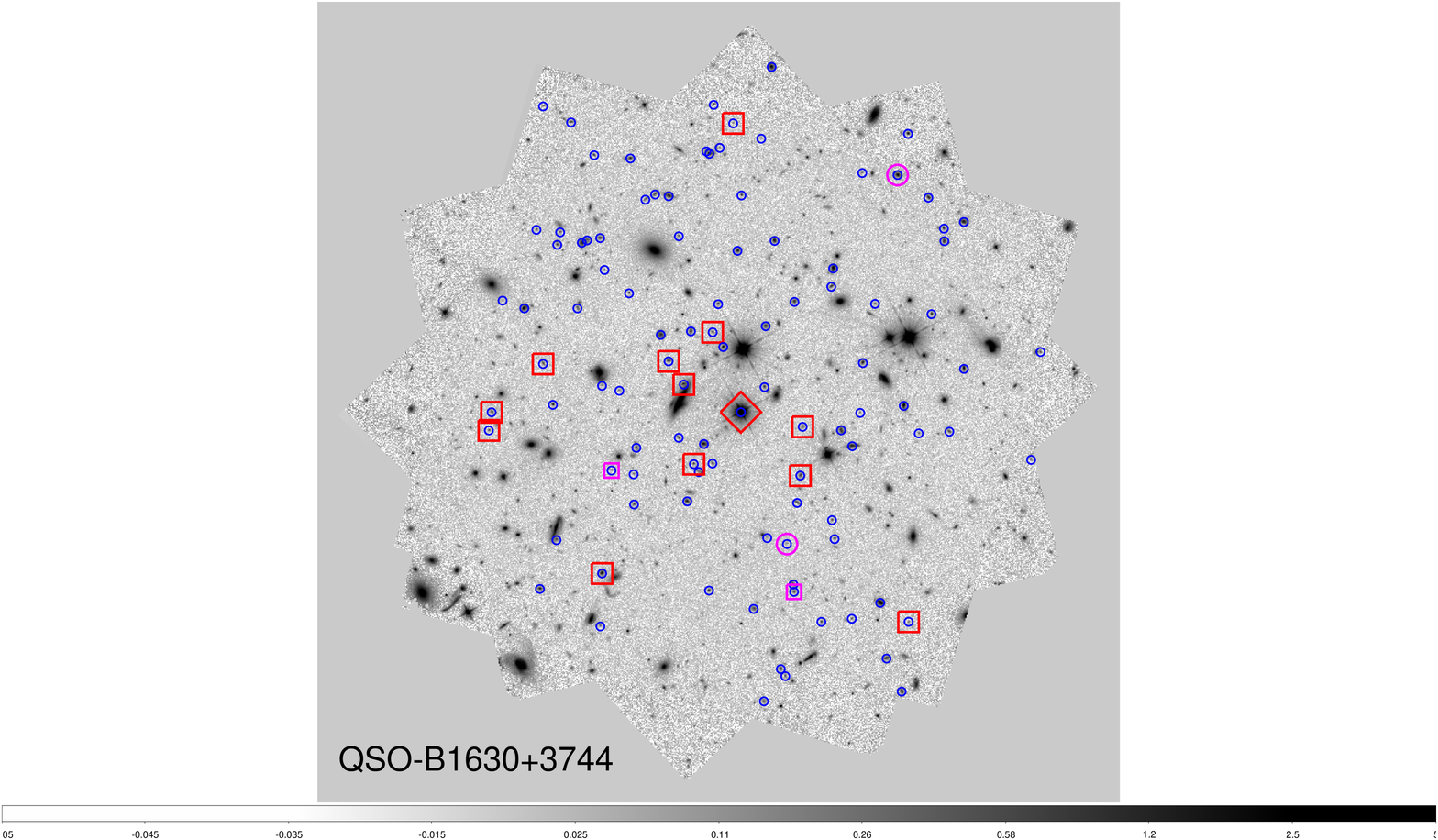}
		\includegraphics[width=0.64\columnwidth, viewport=392 51 1363 1056, clip=true]{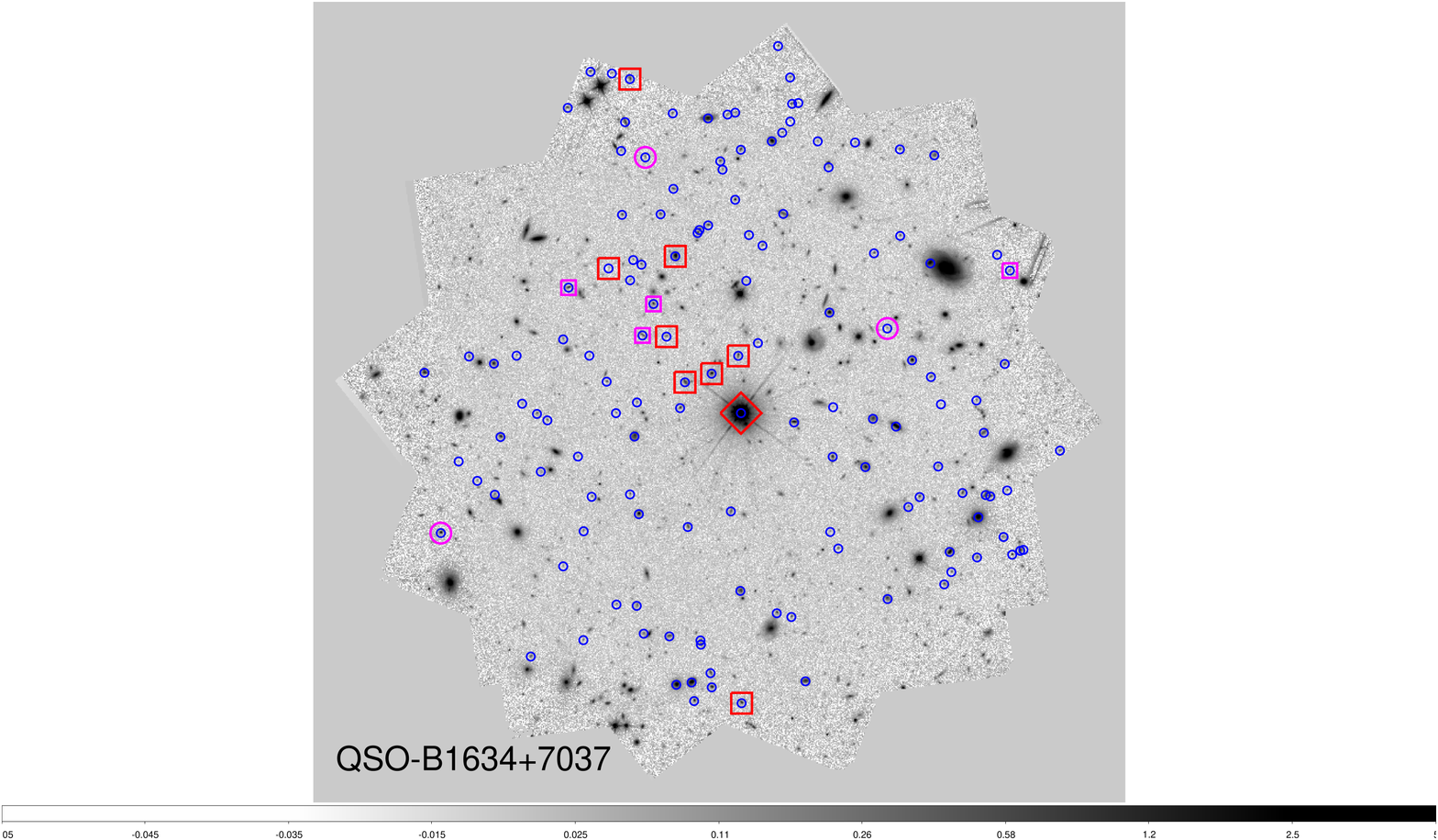}
		
		    \caption[]{The F140W images of the 12 quasar fields. The red diamonds highlight the quasars themselves; the red squares are the overdensities' unambiguous-$z$ members; the magenta circles are the single-line emitting galaxies that would be in the same redshift bin as the quasars if they are either H$\alpha$, [OIII] or [OII] and are therefore likely members; the small magenta squares represent unambiguous-$z$ galaxies in the adjacent $\Delta z=0.025$ bins either side of the quasar bin to highlight any structures that may straddle the bins; and the small blue circles are galaxies in the field with good spectra (quality flags 3 and 4, see \S\ref{sec:spec}). }
	    \label{fig:images}
\end{figure*}

The average background redshift distribution is generated by averaging {all 12 of the galaxy redshift distributions for the different fields together. Specifically, the redshift distributions were placed on a common bin scale of $\Delta z=0.025$, with the mean number of galaxies calculated for each bin to create the background distribution. The common bin scale required small shifts of $<\Delta z=0.0125$ as the quasars are fixed at the centre of their redshift bin for a particular field.} We do not believe that this small shift will cause any biases in the analysis. The bins containing the quasar in each individual field are also removed from the average in case any particularly significant overdensities contaminate the background of the others. For the background value, $N_{\rm bg}$, we take the mean value of this averaged background redshift distribution in the bins with $\Delta z=\pm0.25$ from the quasar, with a $\Delta z=\pm0.05$ buffer around the redshift bin of the quasar being excluded from the analysis in order to avoid an inflated background due to self-contamination. The errors on $\delta_g$ are calculated using the Poisson error on $N_{\rm clust}$ combined in quadrature with the standard error on the background distribution. We note that using the Poisson error means that a galaxy overdensity would need at least 9 members to be considered to have a significance {$\geq3\sigma_\delta$ (where $\sigma_\delta$ is the error on $\delta_g$)}, which may be somewhat conservative {and so we adopt a $>2\sigma_\delta$ threshold, following \cite{galametz2012, wylezalek2013}}. The overdensity values are presented in Table \ref{tab:od}. Of the 12 fields studied, 8 contain $>2\sigma_\delta$ overdensities, showing that $2/3$ of our quasars are in overdense environments. Our richest galaxy overdensity is PG0117+213 with 19 unambiguous-$z$ members and a further 17 potential single-line redshift members, 36 in total (see Fig. \ref{fig:zhistind}). {The additional 17 potential members are single-line galaxies, which would fall into the quasar's $\Delta z=0.025$ redshift bin if the line is confirmed to be H$\alpha$.} The F140W images of the QSAGE fields, with galaxies in the same redshift bin as the quasars highlighted, are displayed in Fig. \ref{fig:images}.

\begin{figure}
		\includegraphics[width=\columnwidth]{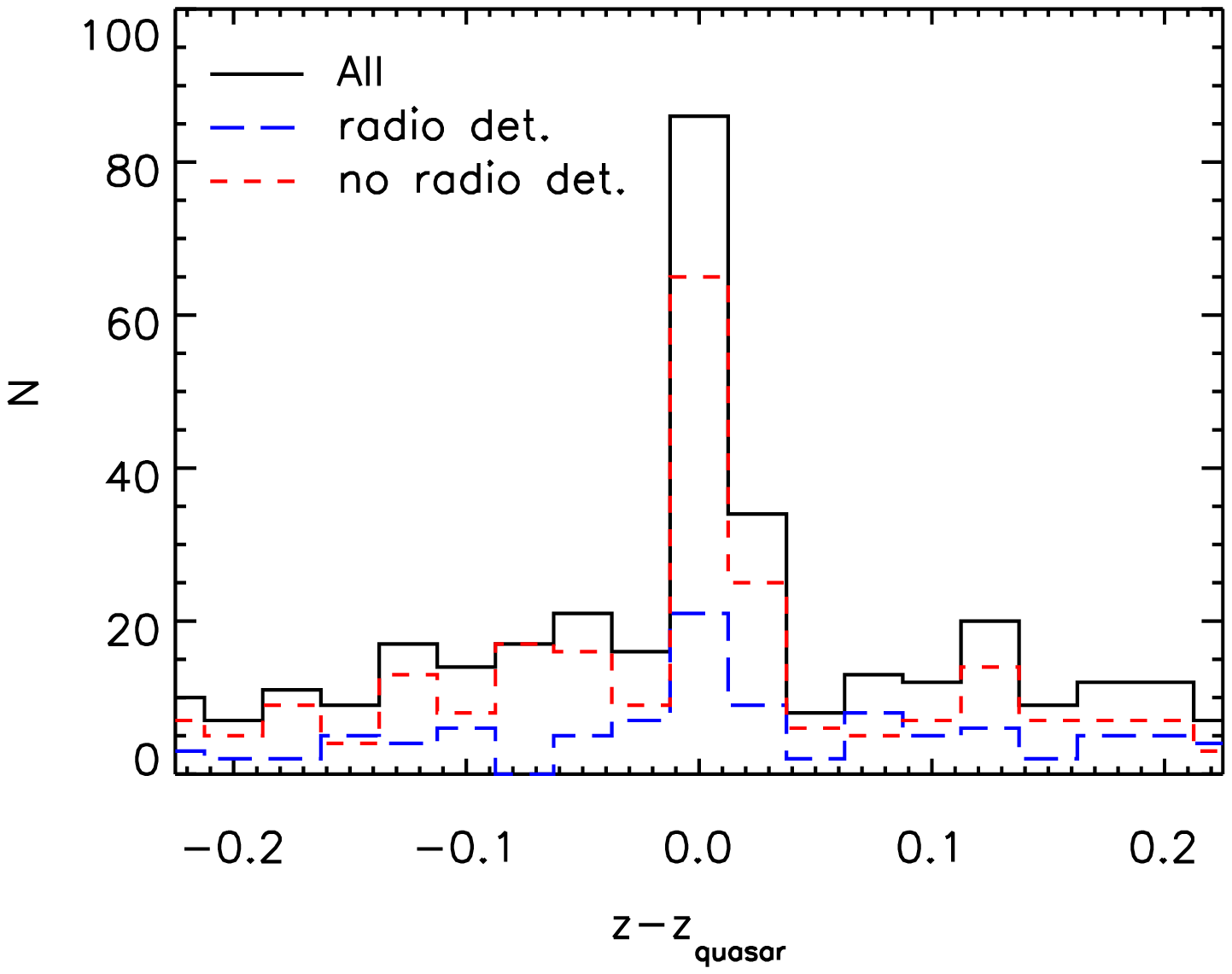}
    \caption{The redshift distributions stacked at the redshift of the individual quasars. The radio and non-radio detections are also stacked independently and show that for this sample radio detected quasars are not more likely to be in overdense environments compared with the general population. }
	    \label{fig:zhiststack}
\end{figure}

In addition to the overdensities calculated for the individual quasar fields, we also stack {the redshift distributions of all 12 fields} in such a way that each field that goes into the stack is centred on the redshift bin of the individual quasars. In this way a stacked redshift `cluster' is created and shown in Fig. \ref{fig:zhiststack}. The overdensity of this stack is $\delta_g=5.8\pm0.9$, {(i.e. a significance of $6.4\,\sigma_\delta$)}, clearly demonstrating that, on average, the QSAGE quasars are in overdense environments.

\subsubsection{Additional overdensities}
We note that there are potentially $\gtrsim2\sigma_\delta$ ($\geq6$ member) overdensities at {redshifts not associated with the quasars} in the fields: HE0515-4414 (6 members); QSO-B0747+4259 (6 members); QSO-B0810+2554 (6, 6 and 7 members); and QSO-B1630+3744 (9 members). These are not analysed further in this paper.

\begin{table*}
	\small
	\centering
	\caption{The number of unambiguous-$z$ members, $N_{\rm clust}$, the number of single-line emitters, $N_{\rm sl}$, and the overdensity value, $\delta_{\rm g}$ and its error, $\sigma_\delta$, for each of the quasar fields. The final row is for the overdensity due to the stack of all 12 fields combined to be centred on the redshift of each quasar (see Fig. \ref{fig:zhiststack}). Also included is a column to indicate whether the quasars are radio emitters, and a column of their absolute $B$-band magnitudes, $M_B$. The R.A. and Dec. columns give the centroid of the overdensity. Finally, $M_{od}$ is the overdensity derived mass. }
	\label{tab:od}
	\begin{tabular}{lcccccccccc} 
		\hline
		field name & $N_{clust}$ & $N_{sl}$ &$\delta_g$ & $\sigma_\delta$ &$\delta_g/\sigma_\delta$ & Radio & $M_B$ & R.A. & Dec. & $\log (M_{od}/M_\odot)$ \\
		\hline

PG0117+213 &           19 &           17 & 17.5 & 4.8 & 3.6 & n & -28.8 & 01:20:17.5 & +21:33:44.6 & 14.7$\pm$0.5\\
PKS-0232-04 &            7 &            4 & 6.3 & 2.9 & 2.2 & y & -28.0 & 02:35:08.3 & -04:01:46.4 & 14.3$\pm$0.4\\
HE0515-4414 &            7 &            1 & 9.7 & 4.4 & 2.2 & n & -29.5 & 05:17:08.8 & -44:10:41.6 & 14.5$\pm$0.4\\
QSO-B0747+4259 &            2 &            3 & 3.5 & 3.3 & 1.1 & n & -29.3 & 07:50:54.4 & +42:52:19.0 & ...\\
QSO-B0810+2554 &            2 &            1 & 1.1 & 1.5 & 0.7 & n & -28.8 & 08:13:30.9 & +25:45:22.9 & ...\\
QSO-J1019+2745 &            7 &            3 & 13.9 & 6.1 & 2.3 & n & -29.3 & 10:19:55.6 & +27:43:57.3 & 14.7$\pm$0.4\\
QSO-B1122-168 &            1 &            0 & ... & ... & ... & y & -28.9 & 11:24:42.9 & -17:05:17.4 & ...\\
QSO-J1130-1449 &            2 &            6 & 1.3 & 1.6 & 0.8 & y & -27.7 & 11:30:05.9 & -14:49:42.4 & ...\\
LBQS-1435-0134 &           11 &            4 & 9.4 & 3.4 & 2.8 & y & -28.9 & 14:37:47.8 & -01:47:12.5 & 14.4$\pm$0.4\\
QSO-B1521+1009 &            6 &            1 & 5.3 & 2.7 & 2.0 & n & -29.4 & 15:24:23.8 & +09:58:29.8 & 14.2$\pm$0.4\\
QSO-B1630+3744 &           13 &            2 & 12.7 & 4.2 & 3.0 & n & -28.3 & 16:32:02.3 & +37:37:49.5 & 14.6$\pm$0.4\\
QSO-B1634+7037 &            9 &            3 & 8.2 & 3.3 & 2.5 & n & -29.2 & 16:34:31.5 & +0:31:47.1 & 14.4$\pm$0.4\\

	\hline
	
	Stack&          86&          45&5.8&0.9&6.6\\

	\hline
	\end{tabular}
\end{table*}

\subsection{Overdensity masses}
\label{sec:mass}

In this section we attempt to estimate the mass of the overdensities in order to understand what kinds of structures they comprise and what they will become by $z=0$. We estimate the masses using the galaxy overdensity parameter, $\delta_g$, following the method of \cite{venemans2005} in which the mass, $M_{od}$ is given by

\begin{equation}
   M_{od}=\bar{\rho} V(1+\delta_m)
	\label{eq:overdense}
\end{equation}

\noindent where $\bar{\rho}$ is the average density of the Universe at the redshift of the galaxy overdensity; $V$ is the comoving volume within which the overdensity is calculated; and $\delta_m$ is the mass overdensity, which is given by  $\delta_m=\delta_g/b$, where $b$ is the bias {of individual galaxy halos and not the overdensity (see e.g. \citealt{steidel1998})}. The comoving volumes probed by our observations are calculated individually with a typical value at $z=1.5$ of $\sim700$\,Mpc$^3$. These overdensities may not be virialised and so $M_{od}$ should be interpreted as the total mass that these systems will collapse to become by $z=0$. To estimate the bias we use the results presented in \cite{cochrane2017}, in which the bias is measured as a function of H$\alpha$ luminosity and redshift. Their relation between bias and $\log L_{\rm H\alpha}$ at $z=1.47$ is the most appropriate to our sample; however the data are noisy at the faint end and do not probe to low enough luminosities for our observations. We therefore fit a relation to the strong linear trend seen between bias and $\log L_{\rm H\alpha}$ at {$z=0.8$} and then {make the assumption that this slope also applies at $z=1.47$. A linear fit is then performed on the $z=1.47$ relation, with the slope fixed to the $z=0.8$ value in order to constrain the $y$ intercept. This is illustrated in Fig. \ref{fig:bias}}. The mean H$\alpha$ luminosity of the galaxies within the overdensities is $\log L_{\rm H\alpha}$\,(erg s$^{-1}$)\,$=41.96$ and so the bias value we estimate is $b=1.45$ based on this analysis of the \cite{cochrane2017} data, which is consistent with the clustering measurements of the QSAGE sample at $z\sim1.4$ (Bielby et al., in prep.). The overdensity derived masses are presented in Table \ref{tab:od}. The mass errors are based solely on the error on $\delta_g$ and not the assumed bias parameter, $b$, and so these errors are potentially larger. If the overdensities are not virialised, then these masses can be considered to be the mass that could potentially collapse out from this volume in order to form a single cluster by $z=0$. The masses derived demonstrate that the mass within each of the 8 significant overdensities is in the range $\log (M/M_\odot)=14.2 - 14.7$, with an average of {$\log (M/M_\odot)=14.5\pm0.2$. We note that due to the size of the, likely underestimated, error bars on the the individual overdensity masses they are all consistent with each other. However, due to lack of a full census of galaxies within the overdensities and the potentially naive assumption made that all mass within the $\sim700$\,Mpc$^3$ volume will collapse into one structure by $z=0$, in reality we expect their final virialised masses to cover a range of values.}

\subsection{Investigating dependencies}
\label{sec:dep}

\begin{figure}
		\includegraphics[width=\columnwidth]{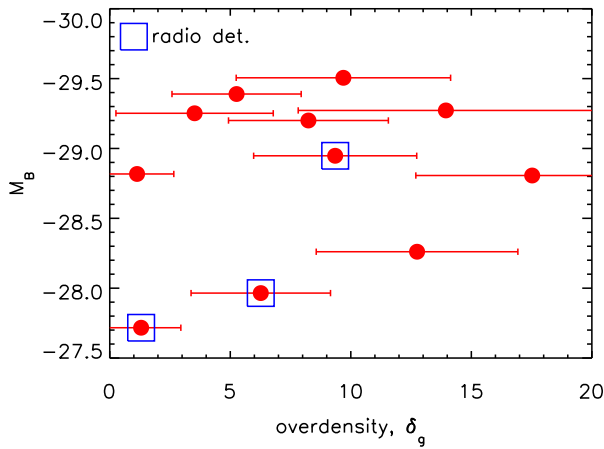}
    \caption[]{The absolute $B$-band magnitude of the quasars, $M_B$, plotted against the overdensity parameter, $\delta_g$, with no correlation seen. }
	    \label{fig:delmb}
\end{figure}

The QSAGE quasars were selected to be luminous in the UV and blue optical to ensure that high signal-to-noise spectra could be obtained for quasar absorption line studies of the intergalactic medium. If quasar luminosity depends on halo mass (e.g. \citealt{silk1998,geach2019}) then we may expect to observe such a correlation. Therefore, we take the $B$-band magnitude from the USNO A2 (\citealt{monet1998}; see Table \ref{tab:sample}) and the quasar $K$ correction from \cite{wisotzki2000} to calculate the absolute $B$-band magnitude in the AB system (assuming a conversion from Vega to AB of $-0.09$ magnitudes). In Fig. \ref{fig:delmb} we plot absolute $B$-band magnitude, $M_B$, against the overdensity parameter, $\delta_g$ and find no correlation. This suggests that while the brightest quasars are often in overdensities, their luminosities do not increase monotonically with $\delta_g$.  

As discussed in \S\ref{sec:intro}, there has been great success in searching for overdense environments in the vicinity of radio galaxies (e.g. \citealt{,venemans2007,hatch2011,wylezalek2013,husband2016}). The quasar sample presented here was not selected to be radio emitting, however some are known radio sources and so we look for a connection between the overdensity parameter and presence of a radio counterpart. All of the QSAGE quasars are covered by the all-sky radio surveys: the National Radio Astronomy Observatory (NRAO) Very Large Array (VLA) Sky Survey (NVSS, \citealt{condon1998}); the Faint Images of the Radio Sky at Twenty-Centimeters (FIRST, \citealt{becker1995}) in the North; and the Sydney University Molonglo Sky Survey (SUMSS, \citealt{mauch2003}) in the South. The coordinates of the QSAGE quasars are entirely covered by a combination of these three surveys and are matched to them using the radii recommended for each survey: NVSS (45''); FIRST (30''); and SUMSS (45''). We find that 4 out of the 12 quasars are radio emitters but see no evidence for greater overdensities in these fields, as shown in Table \ref{tab:od} and Figures \ref{fig:zhiststack} and \ref{fig:delmb}. 

Galaxy clusters and dense environments tend to contain more massive and more passive galaxies than the field population in the local Universe \citep{dressler1980,peng2010}. However, it is expected that at higher redshift there is a reversal of the star formation rate local density relation in order for the constituent galaxies to build up their masses at a faster rate than the field (e.g. \citealt{hilton2010,tran2010}). We therefore compare the properties of the galaxies within the same redshift bin as the quasar to those within the redshift range $1<z<2.5$ {\bf(i.e. the approximate redshift range of the quasar sample)} to see if higher stellar mass and/or more passive galaxies are already in place in these systems or if they are building up their mass at a greater rate. The median stellar mass of all of the {unambiguous$-z$ galaxies (excluding the quasars)} within $1<z<2.5$ is $\log (M/M_\odot) = 9.6 \pm 0.03$ and the median mass of those within the same redshift bin as the quasars {(again excluding the quasars)} is $\log (M/M_\odot)=9.5 \pm 0.09$. The median mass of those in the most significant galaxy overdensity, PG0117+213, is $\log (M/M_\odot)=9.7 \pm 0.2$. There is therefore no evidence for a difference in mass between the population in the same structure as the quasar and that of the field. However, due to the line-emission selection we may be missing a population of high mass passive galaxies.

The median H$\alpha$ derived SFR for the {unambiguous$-z$ galaxies (excluding the quasars)} with redshifts $0.7<z<1.6$ {(i.e. the redshift range over which we can detect H$\alpha$)} is $3.7\pm0.5$\,M$_\odot$yr$^{-1}$ and for those in the same redshift bins as their quasars {(again excluding the quasars)} it is $4.0\pm1.9$\,M$_\odot$yr$^{-1}$. Therefore, there is no evidence for a difference between the SFR of the galaxies within the overdensities and those in the field. {However, as discussed in \S\ref{sec:spec}, a systematic difference in gas phase metallicity between the field and overdensities could affect this, as we assume a single metallicity for the whole sample to extract H$\alpha$ flux from the blended H$\alpha$ and [NII] line. Additionally}, due to the line-emission selection we may be missing a population of more passive galaxies.

\section{Discussion}
\label{sec:disc}

From the analyses of the redshift distributions of the individual quasar fields, we find that 8/12 of the QSAGE quasars are in overdensities and the combined stack of the redshift distributions confirms this is true on average (see Fig. \ref{fig:zhiststack}). The overdensities have inferred masses of $\log (M/M_\odot)=14.2-14.7$, indicating that they are either already galaxy clusters or regions of space that will collapse further into galaxy clusters by $z=0$. The richest galaxy overdensity in our sample is found around the quasar PG0117+213, which has 19 confirmed members and potentially 36 in total and as such, its predicted mass of $\log (M/M_\odot)=14.7$ is likely a lower limit. We note that a narrow-band survey would have determined that all line-emitters within the narrow-band filter wavelength range were potential members. One of the least dense fields is QSO-B0810+2554, which is the lensed quasar discussed in \S\ref{sec:qso}, there is a possibility that this quasar's intrinsic brightness has been boosted by the lensing to bring it into the sample. In general, the overdensity values and masses will be lower limits, as we did not include the single-line emitters with ambiguous redshifts and our emission line selection will not have detected the passive galaxies present in $z=1-2$ clusters. 

Overdensities such as these have often been found for radio galaxy fields (e.g. \citealt{,venemans2007,hatch2011,wylezalek2013,husband2016}) but this has been less clear for those surrounding quasars, which typically reside in $\log (M/M_\odot)=12 - 13$ dark matter halos \citep{kauffmann2008,hickox2009,ross2009,shen2009,geach2019}. However, as discussed in \S\ref{sec:intro} there are both theoretical arguments and observational hints that the most UV/optically bright quasars live in higher mass halos. The feedback regulated growth of black holes should result in a relation between black hole mass and halo mass e.g. $M_{bh}\propto M_h^{5/3}$ (\citealt{silk1998,wyithe2003}) and so assuming accretion is approximately at a fixed fraction of the Eddington limit, the quasar luminosity $L\propto M_h^{5/3}$. Statistical studies have attempted to confirm this, finding that the brighter quasars in their clustering analyses reside in higher mass halos (e.g. $M_h \approx10^{13} M_\odot h^{-1}$, \citealt{geach2019}). However, results at $z=4$ show no evidence that the brightest quasars live in overdensities \citep{uchiyama2018}. 

Our QSAGE sample includes some of the brightest quasars at $z=1-2$, and we have demonstrated, for the first time, that the brightest quasars really are in highly overdense environments. We speculate that the reason this is not seen in lower redshift clusters is that clusters in the local Universe tend to be dominated by passive red sequence galaxies \citep{dressler1980,peng2010}, which contain significantly less cold gas to fuel a quasar. At $z>1$, there must be a reversal of this relation in order for the cluster galaxies to build up their stellar mass through star formation and so clusters and protoclusters contain more gas rich galaxies at early times (e.g. \citealt{hilton2010,tran2010}). This combination of a high halo mass, gas-rich environment means that luminous quasars are more likely to be good tracers of $z>1$ cluster and protoclusters rather than their low redshift counterparts. Our results demonstrate that searching for overdensities around the brightest quasars is an excellent way to find high redshift clusters and protoclusters. 

The discrepancy between our results and the lack of overdensities found around bright quasars at $z=4$ \citep{uchiyama2018}, may be due to several factors. Firstly, our redshift range is lower and so perhaps the bright quasar population live in low mass halos at $z\sim4$, with little evolutionary connection to their $z=1-2$ counterparts. Secondly, the \cite{uchiyama2018} study uses a photometric overdensity catalogue created from $g$-band drop out galaxies, whereas our deep grism data means that we obtain a spectroscopic redshift census of the line-emitting galaxies in the field. It may therefore be the case that we can probe structures that would be missed by a photometric technique.

One potential reason for finding the brightest quasars within overdense environments is that they may be regions of high merger or interaction activity, which will both increase the black hole's mass through merging and accretion, as gas is disturbed and funnelled towards the black hole. Another related explanation is that these luminous quasars contain the most massive black holes (and therefore have high Eddington luminosities) in the cluster because they are the progenitors of the BCG population, going through a rapid stellar and black hole mass growth phase via mergers and/or accretion \citep{hlavacek2013}. These scenarios are in agreement with the `blowout' phase of the evolutionary path described by \cite{hopkins2008}.

Within the sample, we find there is no relationship between the luminosity of the quasar and the strength of the overdensity. This may in part be related to the variable nature of quasar luminosity (e.g. \citealt{macleod2012}), which is governed by variability in the fuelling rate, and/or as the observations are rest-frame UV, a small amount of dust obscuration could also act to remove any correlation. 

There is no evidence for enhancement of the overdensities around the radio emitting quasars compared with the sample as a whole. This is in apparent contrast to \cite{wylezalek2013}, although only $\sim50\%$ of their sample of radio-loud quasars show such an enhancement and our survey contains only four radio emitters so we do not draw any conclusions from this. 

Finally, there is no evidence that the galaxy populations within the overdensities are different to those in the field in terms of stellar mass or SFR. This is further evidence for a reversal of SFR - local density relation at $z>1$ as the overdensity members appear to be forming stars at the same rate as their field counterparts, in contrast to $z<1$ clusters. However, as this sample is selected on line-emission, we may be missing a significant number of massive and passive galaxies. 

\section{Summary \& conclusions}
\label{sec:conc}

In this paper, we demonstrate that 8/12 of the QSAGE quasars reside in overdensities with masses of $\log (M/M_\odot)=14.2-14.7$. This is strong evidence to support the hypothesis that the brightest quasars should live in the most massive dark matter halos (\citealt{silk1998,wyithe2003}). It also paves the way for future narrow-band, grism or integrated field spectroscopy studies of the populations around high-$z$ quasars in order to look for protoclusters. Of particular note is PG0117+213, which is a rich structure with up to 36 members within a cylinder of $\sim700$\,kpc radius and $\delta z=0.025$. We speculate that the reason we are finding bright quasars within these overdensities is due to a increased mergers activity and/or even the mass build up of a BCG progenitor. The galaxies within the overdensities appear to have similar masses and SFR to those in the field, which is further evidence for a reversal of SFR - local density relation at $z>1$.

This paper represents part of the galaxy evolution component of the QSAGE survey, which has the primary goal of obtaining hundreds of galaxy redshifts in order to associate absorption features in the UV spectra of quasars with their host galaxies' CGM. Combined, these two components of the QSAGE survey will provide comprehensive constraints to the galaxy fuelling and feedback process in galaxy formation models.(e.g. \citealt{bielby2019} and forthcoming papers).

\section*{Acknowledgements}

{We first thank the anonymous referee for improving the clarity of the paper. We also} thank James Mullaney, Nic Ross and David Sobral for useful discussions. RMB, MF, RGB, and SLM acknowledge the Science and Technology Facilities Council (STFC) through grants ST/P000541/1 and ST/L00075X/1 for support. This project has received funding from the European Research Council (ERC) under the European Union's Horizon 2020 research and innovation programme (grant agreement No 757535). This work has been supported by Fondazione Cariplo, grant number 2018-2329. JNB and JXP received financial support for this research through NASA Grant HST-GO-11741 from the Space Telescope Science Institute, which is operated by the Association of Universities for Research in Astronomy, Inc., under NASA Contract NAS5-26555. RAC is a Royal Society University Research Fellow.

\section*{Data availability}

The data underlying this article are from HST Cycle 24 proposal 14594: `QSAGE: QSO Sightline And Galaxy Evolution' and are publicly available from the Mikulski Archive for Space Telescopes (MAST, https://archive.stsci.edu/hst/).




\bibliographystyle{mnras}
\bibliography{QSOclust_finalmnras} 




\appendix

\section{Additional Material}

\begin{figure}
		
    	 \includegraphics[width=\columnwidth]{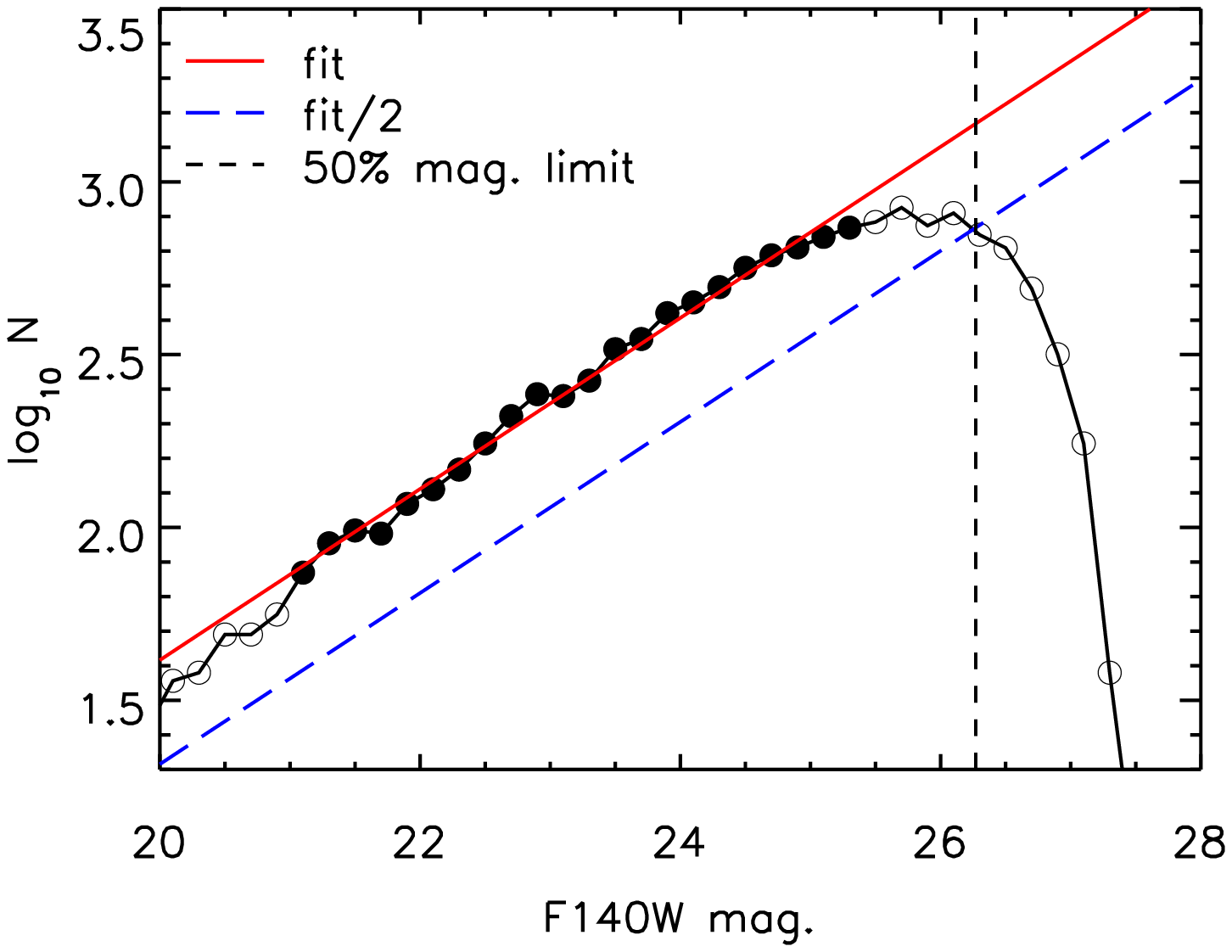}

	     \caption{{A plot of the logarithm of the F140W detection magnitude distribution to demonstrate how the 50\% completeness limit for F140W magnitude was approximated. The solid red line is a fit to the approximately linear portion of the plot (filled black points). The long dashed blue line is offset from the red line by a factor of 2 (i.e. the 50\% line). The vertical dashed line marks where the data crosses the 50\% line at 26.3 mag, which is the approximate 50\% completeness limit.}}

 \label{fig:maglim}
\end{figure}

\begin{figure*}
		    	 \includegraphics[width=0.66\columnwidth]{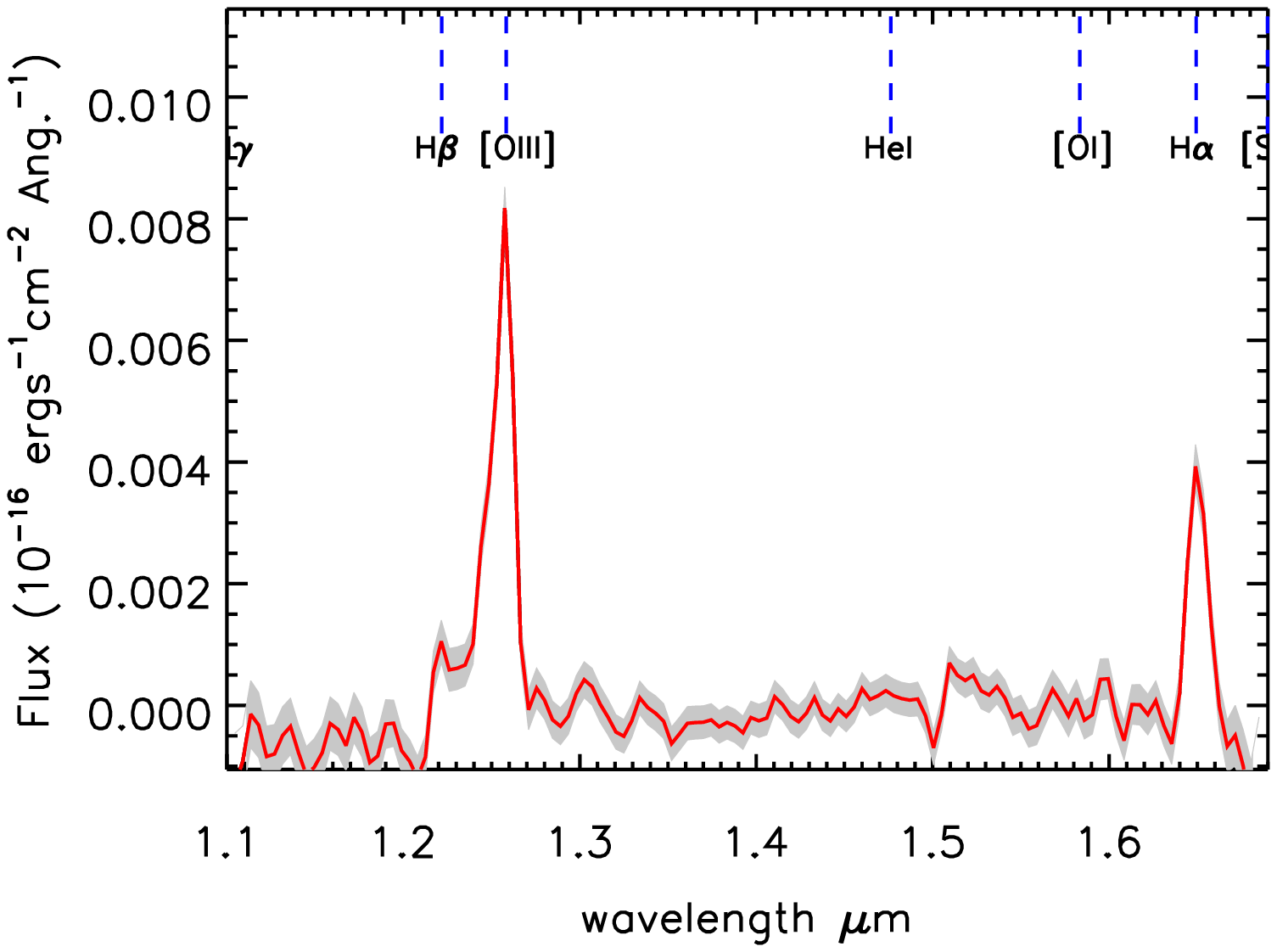}
			   \includegraphics[width=0.66\columnwidth]{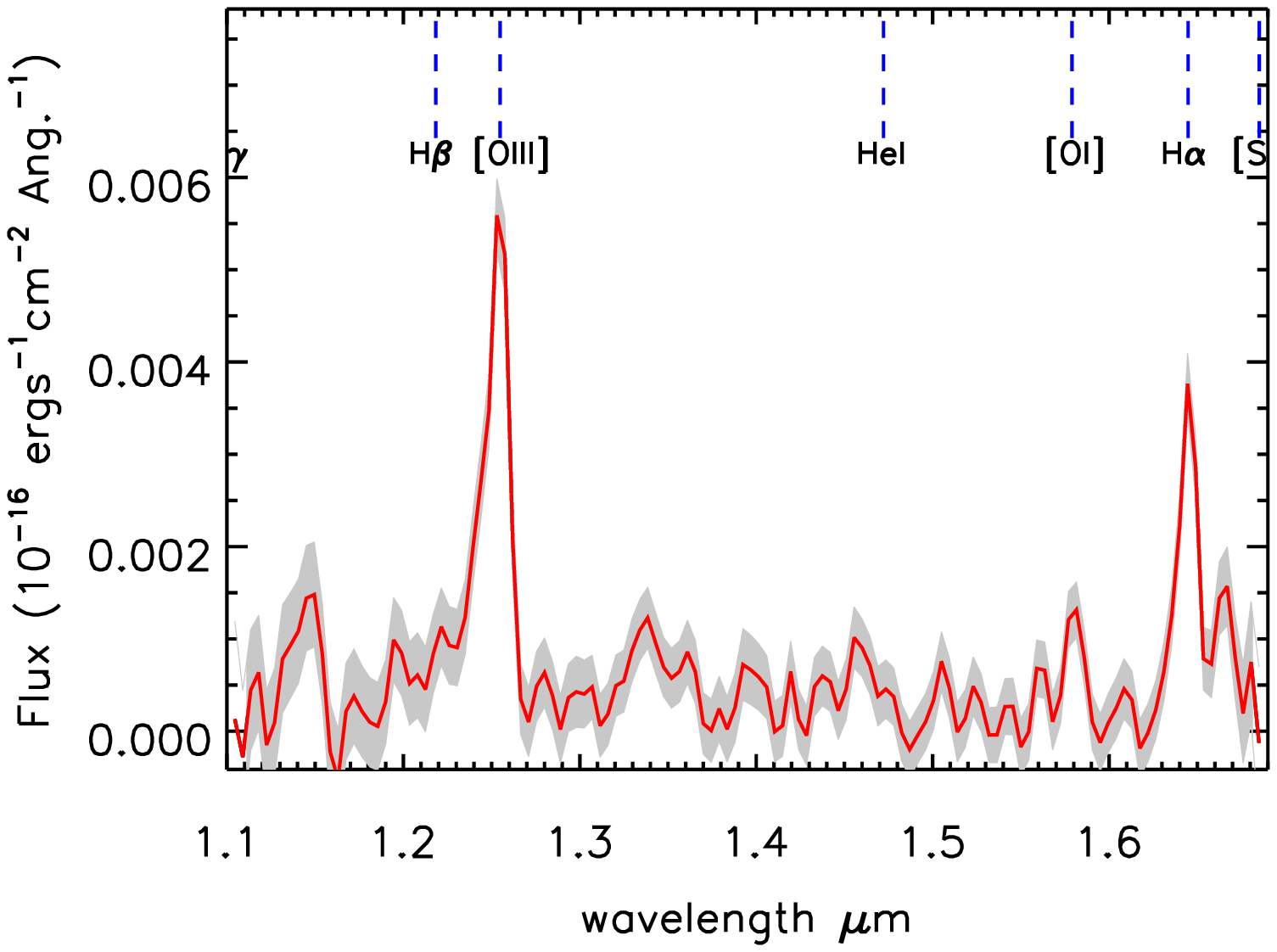}
  			  \includegraphics[width=0.66\columnwidth]{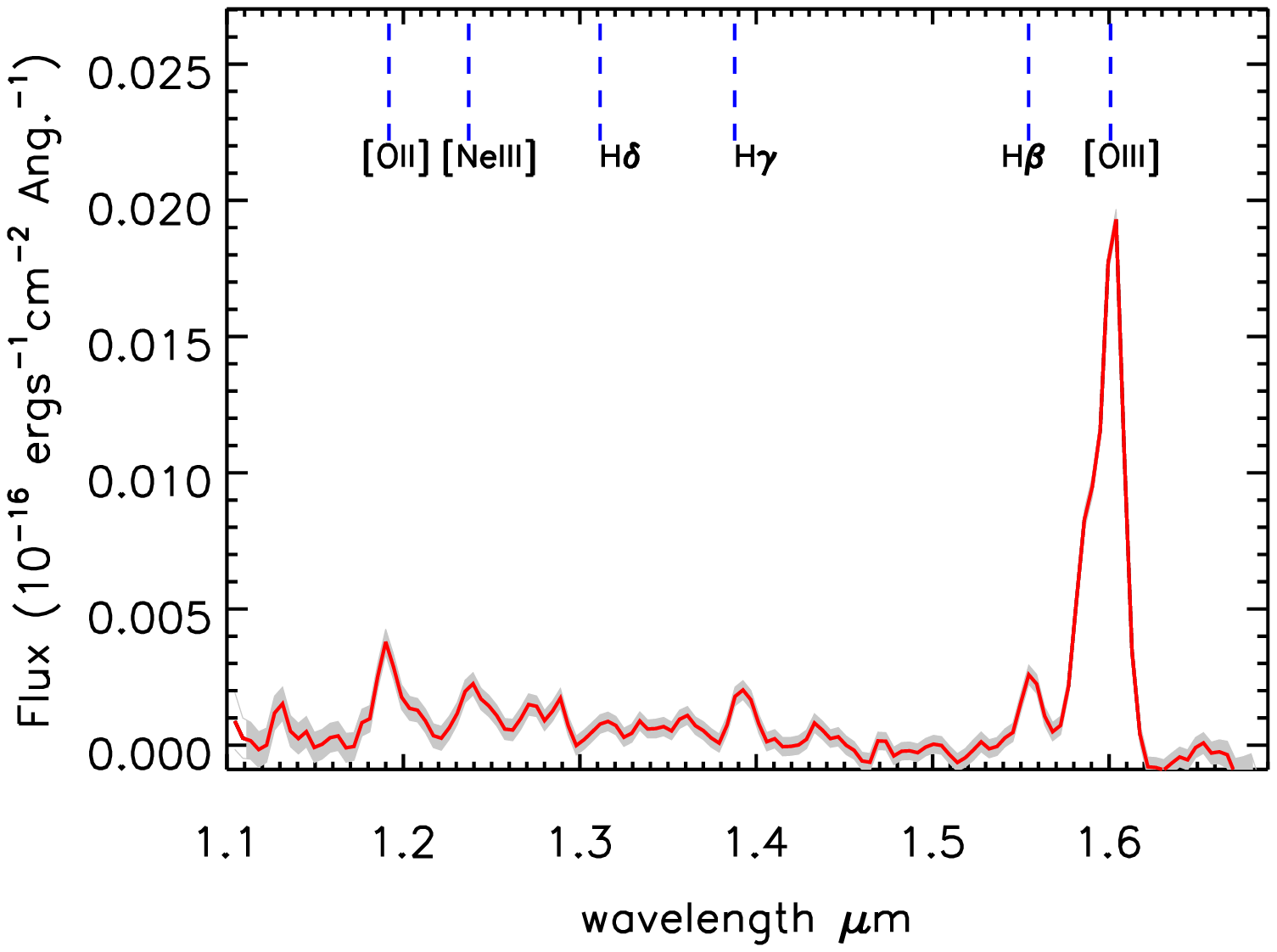}
			  \includegraphics[width=0.66\columnwidth]{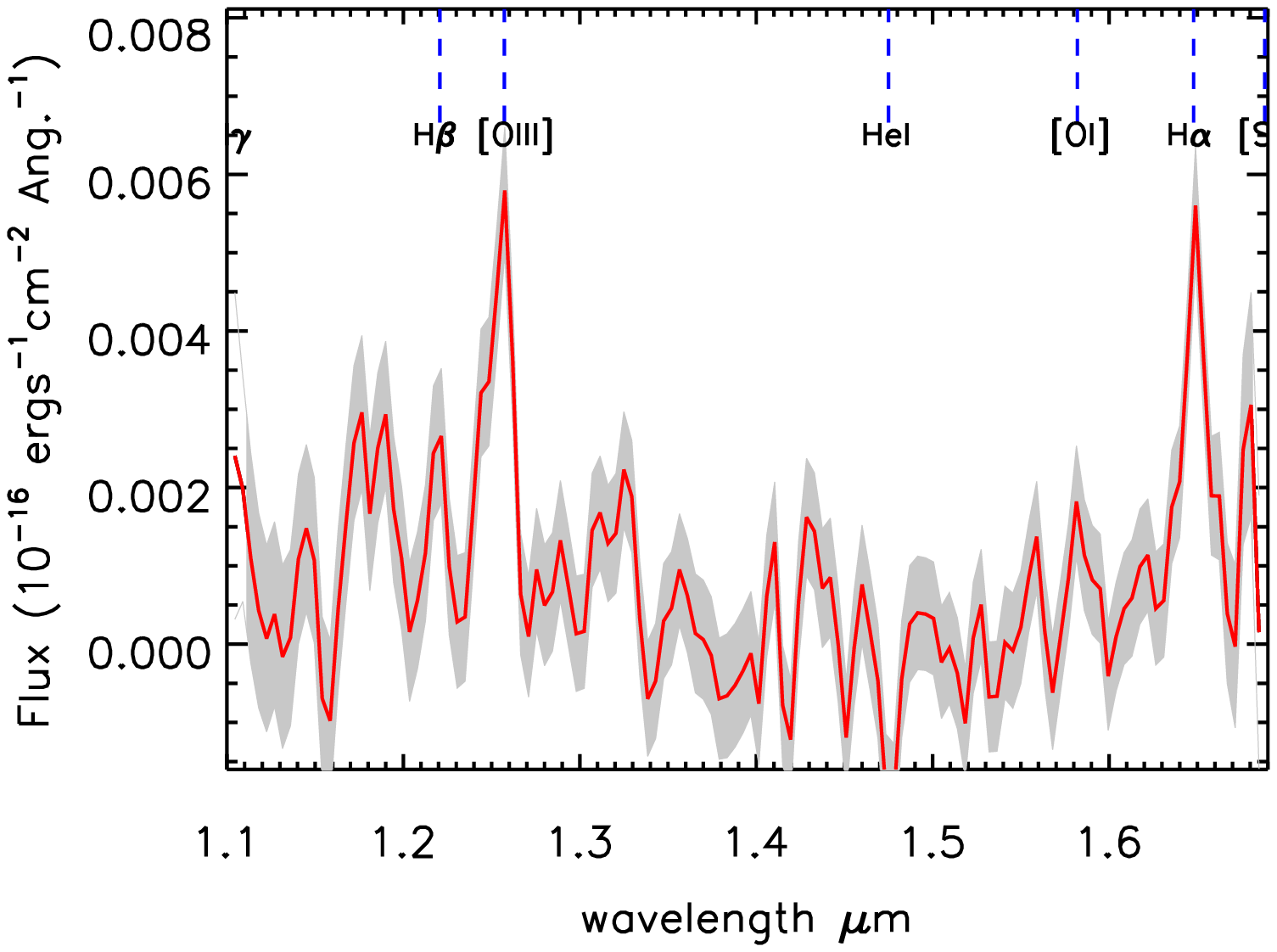}
			\includegraphics[width=0.66\columnwidth]{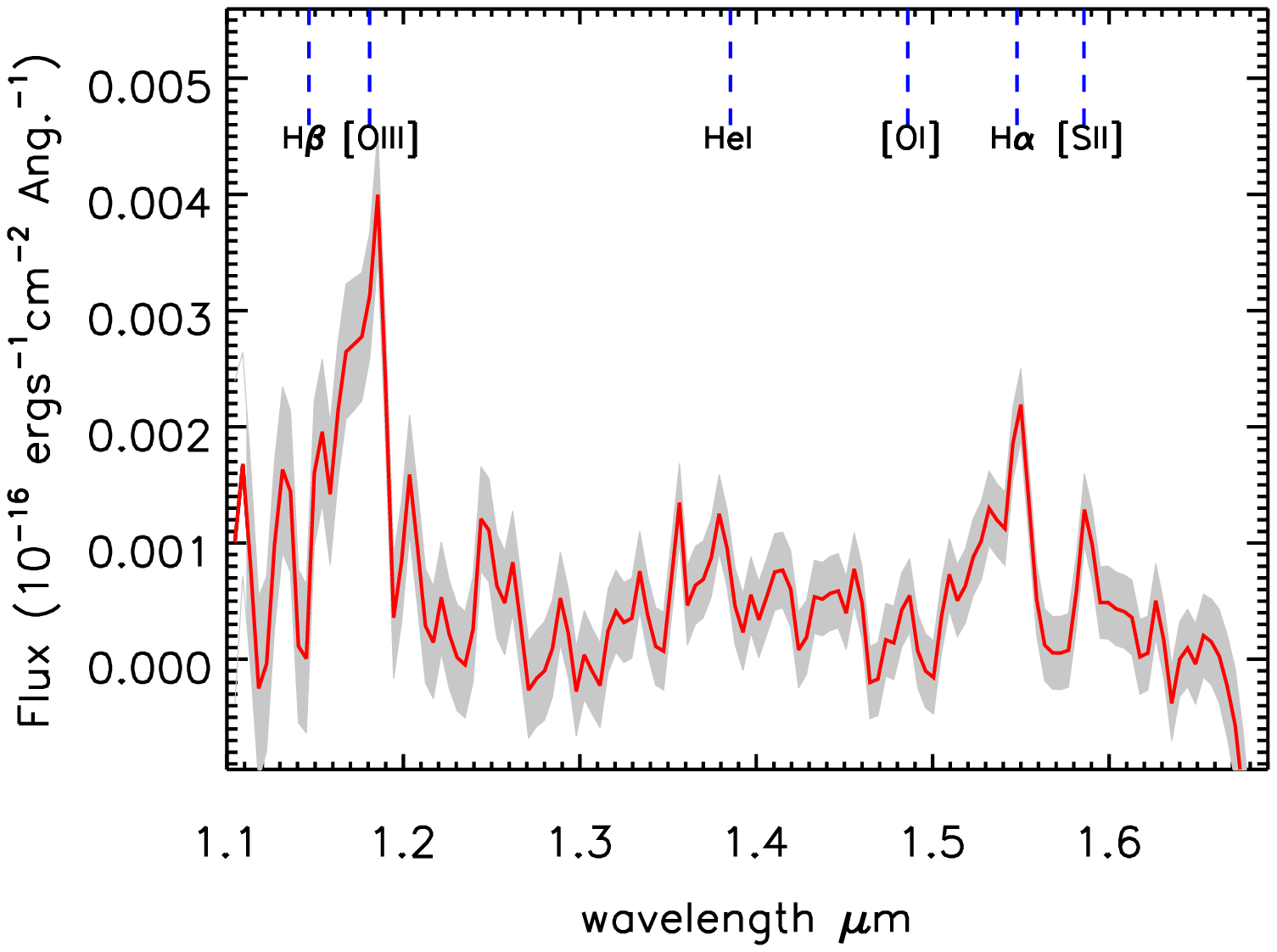}
			\includegraphics[width=0.66\columnwidth]{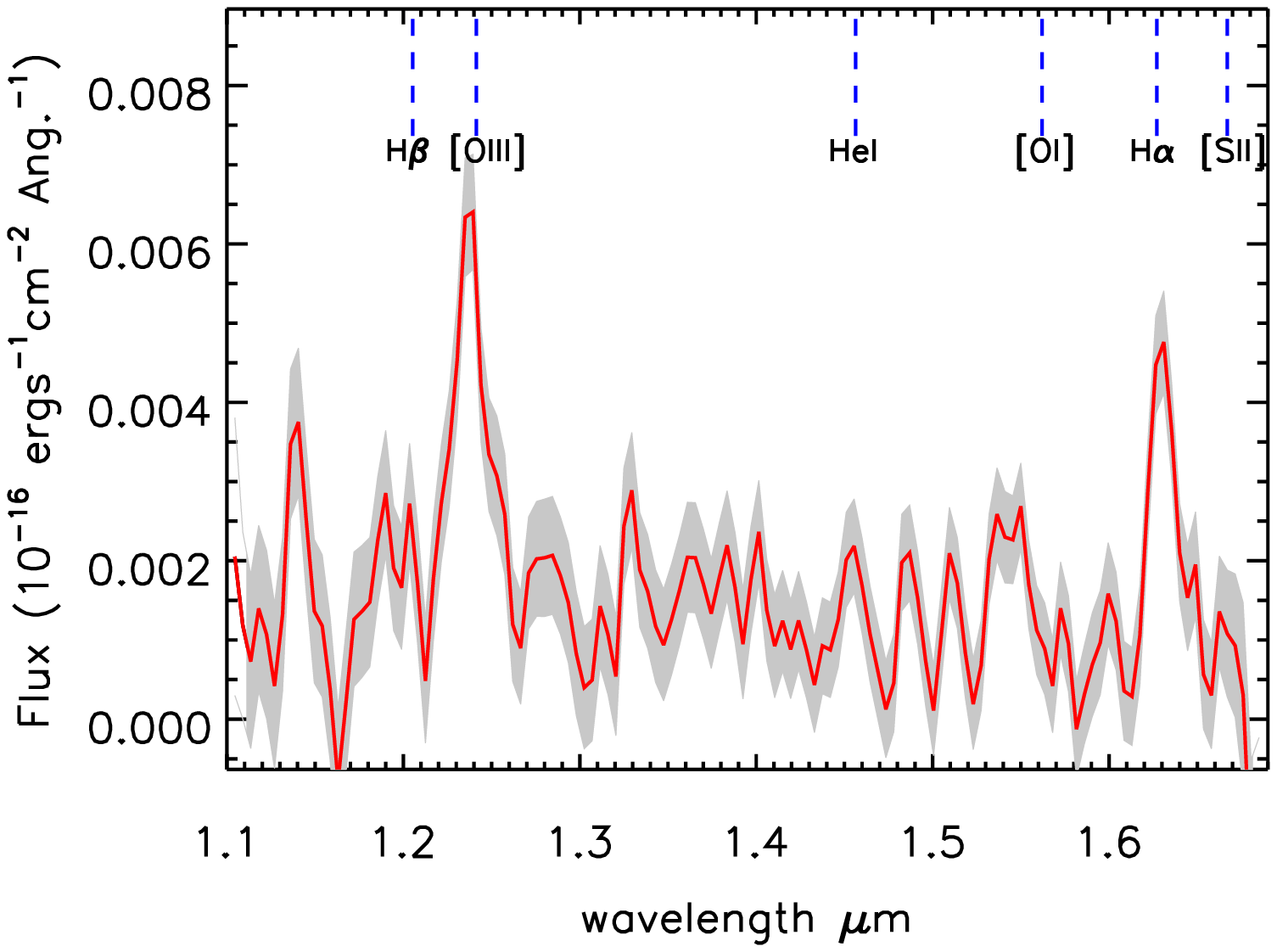}

	     \caption{{Example WFC3 spectra from the sample used in our analysis. The upper row are quality flag 4 spectra (line S/N$>10$) and the lower row are the quality flag 3 spectra (line S/N$>3$)}}

 \label{fig:exspec}
\end{figure*}

\begin{table}
	\footnotesize
	\centering
	\caption[]{The SED stellar mass to F160W magnitude calibration derived from the CANDELS catalogues of \cite{barro2019}. This is a simple linear fit of the form $\log (M_{*}/M_\odot)=a\,m_{\rm F160W} + m_{\rm intercept}$. An example fit is given in Fig. \ref{fig:masscal}}
	\label{tab:massfit}
	\begin{tabular}{lcc} 
		\hline
		$z$ & $m_{\rm intercept}$ & $a$  \\
		\hline
$0.3-0.4$ & 18.95$\pm$0.14 & -0.4542$\pm$0.005\\
$0.4-0.5$ & 18.91$\pm$0.08 & -0.4428$\pm$0.003\\
$0.5-0.6$ & 19.26$\pm$0.08 & -0.4511$\pm$0.003\\
$0.6-0.7$ & 19.57$\pm$0.11 & -0.4558$\pm$0.004\\
$0.7-0.8$ & 19.63$\pm$0.09 & -0.4520$\pm$0.004\\
$0.8-0.9$ & 20.02$\pm$0.09 & -0.4636$\pm$0.004\\
$0.9-1.0$ & 20.11$\pm$0.08 & -0.4639$\pm$0.003\\
$1.0-1.1$ & 20.33$\pm$0.10 & -0.4701$\pm$0.004\\
$1.1-1.2$ & 20.94$\pm$0.13 & -0.4925$\pm$0.005\\
$1.2-1.3$ & 20.94$\pm$0.11 & -0.4900$\pm$0.004\\
$1.3-1.4$ & 21.39$\pm$0.12 & -0.5045$\pm$0.005\\
$1.4-1.5$ & 21.62$\pm$0.15 & -0.5120$\pm$0.006\\
$1.5-1.6$ & 21.74$\pm$0.14 & -0.5147$\pm$0.006\\
$1.6-1.7$ & 21.79$\pm$0.15 & -0.5134$\pm$0.006\\
$1.7-1.8$ & 21.55$\pm$0.19 & -0.5009$\pm$0.007\\
$1.8-1.9$ & 21.48$\pm$0.23 & -0.4971$\pm$0.009\\
$1.9-2.0$ & 21.32$\pm$0.21 & -0.4892$\pm$0.008\\
$2.0-2.1$ & 21.60$\pm$0.23 & -0.4968$\pm$0.009\\
$2.1-2.2$ & 21.03$\pm$0.25 & -0.4708$\pm$0.010\\
$2.2-2.3$ & 21.03$\pm$0.25 & -0.4708$\pm$0.010\\
$2.3-2.4$ & 20.98$\pm$0.28 & -0.4666$\pm$0.011\\
$2.4-2.5$ & 20.62$\pm$0.30 & -0.4515$\pm$0.012\\
$2.5-2.6$ & 20.84$\pm$0.36 & -0.4616$\pm$0.014\\
$2.6-2.7$ & 20.35$\pm$0.42 & -0.4421$\pm$0.016\\
$2.7-2.8$ & 20.35$\pm$0.46 & -0.4407$\pm$0.018\\
$2.8-2.9$ & 20.80$\pm$0.49 & -0.4535$\pm$0.019\\
$2.9-3.0$ & 19.67$\pm$0.37 & -0.4066$\pm$0.014\\
$3.0-3.1$ & 18.74$\pm$0.44 & -0.3697$\pm$0.017\\
$3.1-3.2$ & 18.91$\pm$0.45 & -0.3717$\pm$0.017\\
$3.2-3.3$ & 18.39$\pm$0.48 & -0.3496$\pm$0.019\\
		\hline
	\end{tabular}
\end{table}

\begin{figure}
		\includegraphics[width=\columnwidth]{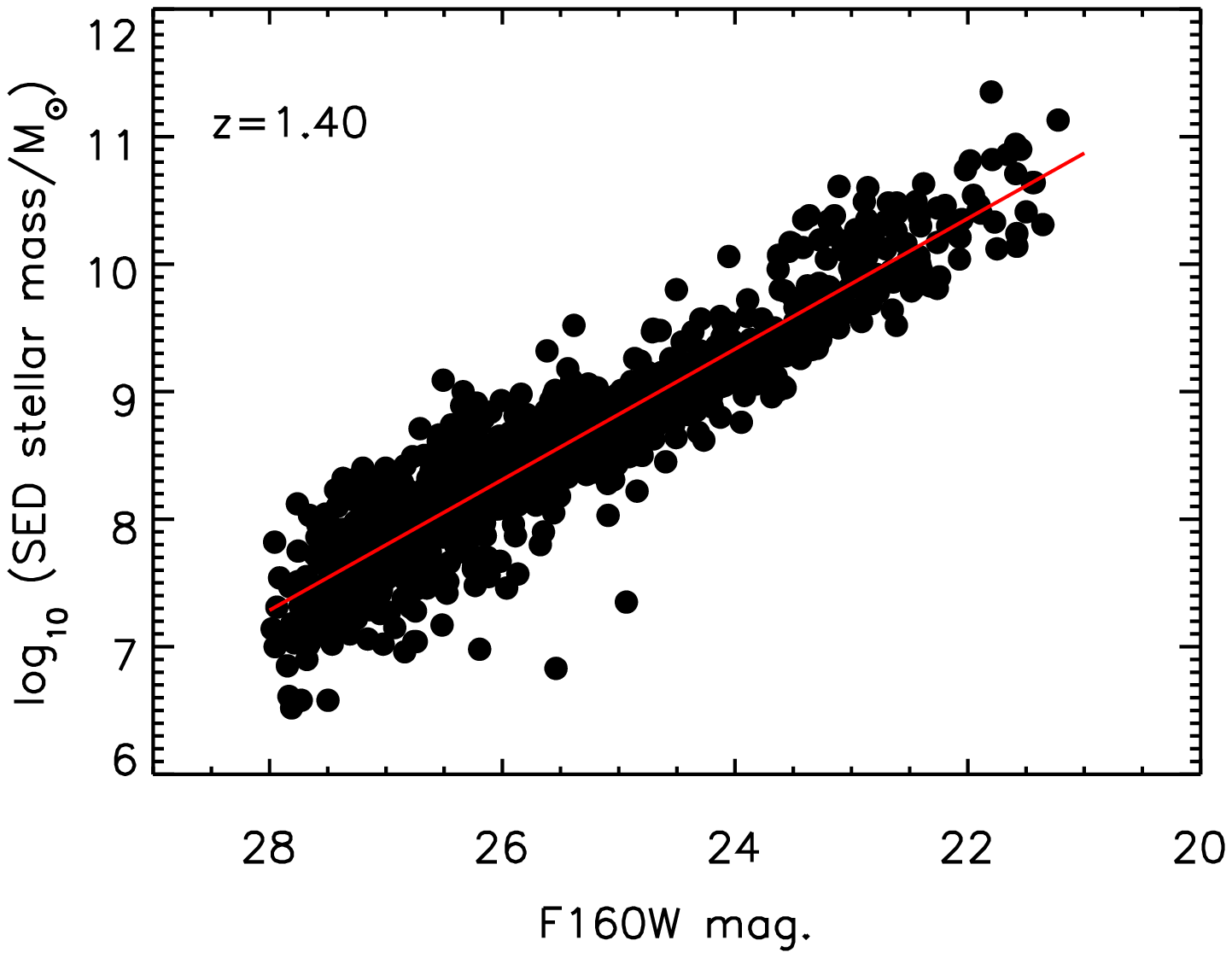}
    \caption[]{The $z=1.5$ calibration of SED-fit stellar mass to WFC3 F160W magnitude using the CANDELS catalogues of \cite{barro2019}. Table \ref{tab:massfit} contains the fit parameters for all redshifts.}
	    \label{fig:masscal}
\end{figure}

\begin{figure}
		\includegraphics[width=\columnwidth]{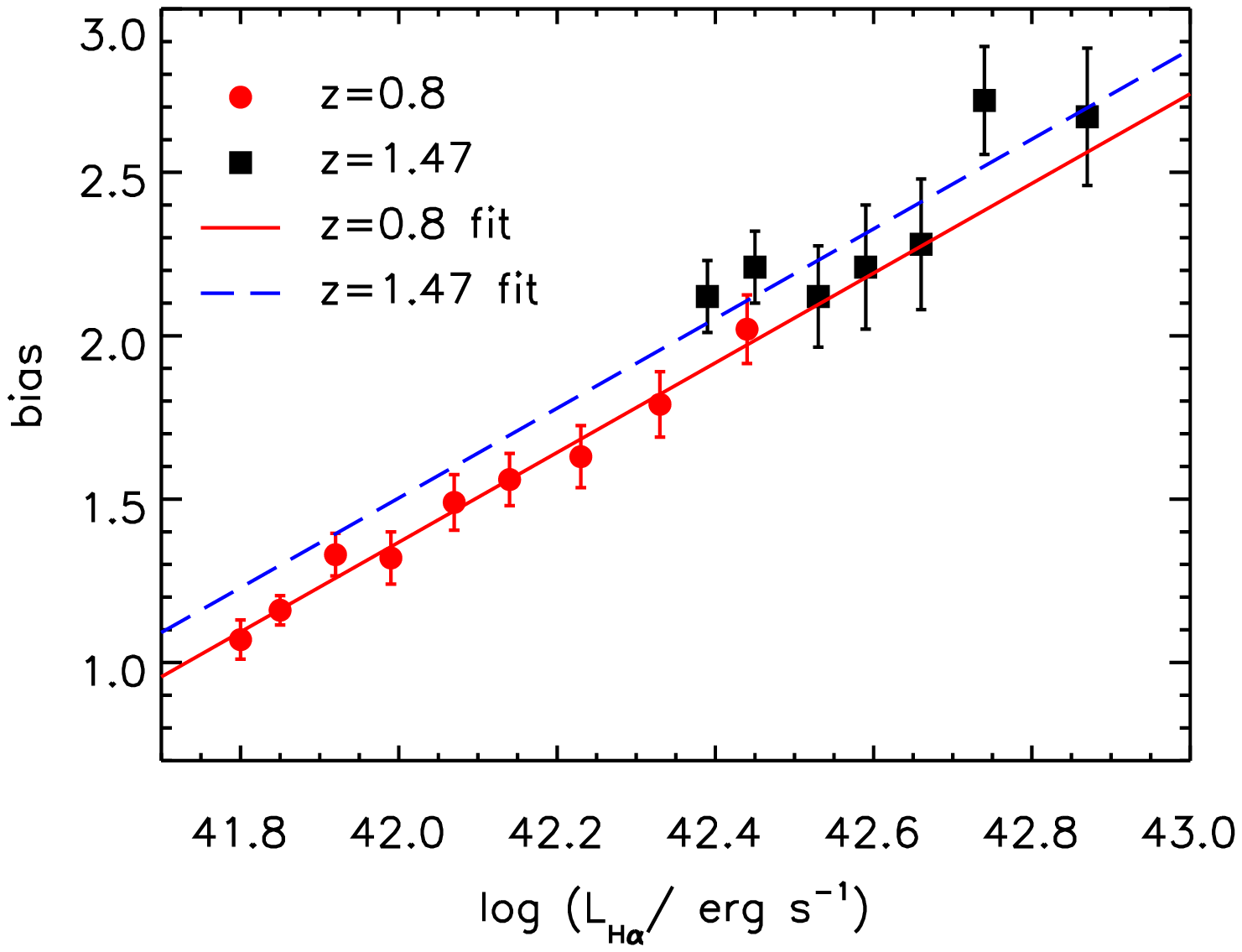}
    \caption[]{{The bias parameter, $b$, value as a function of H$\alpha$ luminosity for H$\alpha$ emitting galaxies at $z=0.8$ and z=$1.47$ from \cite{cochrane2017}. The red solid line is a fit to the $z=0.8$ data and the blue dashed line is a fit to the $z=1.47$, with the slope fixed to that of the $z=0.8$ fit. This was used to estimate the bias parameter, $b$, for the galaxies in our sample. The average redshift of our sample is $z=1.5$  and the average H$\alpha$ luminosity of the galaxies within overdensities is $\log L_{\rm H\alpha}$\,(erg s$^{-1}$)\,$=41.96$ and so we estimate that $b=1.45$. }}
	    \label{fig:bias}
\end{figure}



\bsp	
\label{lastpage}
\end{document}